\pdfoutput=1
\documentclass[twocolumn]{aastex631}
\usepackage{graphicx}
\usepackage{amsmath} 
\usepackage{xcolor}
\usepackage{subfigure}
\usepackage{tikz}
\tikzset{
  load/.style   = {ultra thick,-latex},
  stress/.style = {-latex},
  dim/.style    = {latex-latex},
  axis/.style   = {-latex},
}
\usepackage{hyperref}

\newcommand{\vb}[1]{\boldsymbol{#1}}
\newcommand{\nablab}{\boldsymbol{\nabla}}

\newcommand{\dpartial}[2]{\frac{\partial #1}{\partial #2}}

\newcommand{\dtotal}[2]{\frac{\mathrm{d} #1}{\mathrm{d} #2}}
\newcommand{\ddtotal}[2]{\frac{\mathrm{d}^2 #1}{\mathrm{d} #2^2}}

\newcommand{\tw}{\the\textwidth}
\newcommand{\cw}{\the\columnwidth}
\graphicspath{{./}{figures/}}

\submitjournal{ApJ}
\received{Dec 6, 2022}
\revised{Mar 22, 2023}
\accepted{Apr 13, 2023}
\shorttitle{Tides in stratified planets}
\shortauthors{Pontin, Barker \& Hollerbach}

\begin{document}
\title{Tidal dissipation in stratified and semi-convective regions of giant planets}

\author{Christina M. Pontin}
\affiliation{Department of Applied Mathematics, School of Mathematics, University of Leeds, Leeds, LS2 9JT, UK}

\author[0000-0003-4397-7332]{Adrian J. Barker}
\affiliation{Department of Applied Mathematics, School of Mathematics, University of Leeds, Leeds, LS2 9JT, UK}
\correspondingauthor{Adrian Barker}
\email{A.J.Barker@leeds.ac.uk}

\author[0000-0001-8639-0967]{Rainer Hollerbach}
\affiliation{Department of Applied Mathematics, School of Mathematics, University of Leeds, Leeds, LS2 9JT, UK}
\email{R.Hollerbach@leeds.ac.uk}

\begin{abstract}
We study how stably stratified or semi-convective layers alter the tidal dissipation rates associated with the generation of internal waves in planetary interiors. We consider if these layers could contribute to the high rates of tidal dissipation observed for Jupiter and Saturn in our solar system. We use an idealised global spherical Boussinesq model to study the influence of stable stratification and semi-convective layers on tidal dissipation rates. We carry out analytical and numerical calculations considering realistic tidal forcing and measure how the viscous and thermal dissipation rates depend on the parameters relating to the internal stratification profile. We find that the strongly frequency-dependent tidal dissipation rate is highly dependent on the parameters relating to the stable stratification, with strong resonant peaks that align with the internal modes of the system. The locations and sizes of these resonances depend on the form and parameters of the stratification, which we explore both analytically and numerically. Our results suggest that stable stratification can significantly enhance the tidal dissipation in particular frequency ranges. Analytical calculations in the low frequency regime give us scaling laws for the key parameters, including the tidal quality factor $Q'$ due to internal gravity waves. Stably stratified layers can significantly contribute to  tidal dissipation in solar and extrasolar giant planets, and we estimate substantial tidal evolution for hot Neptunes. Further investigation is needed to robustly quantify the significance of the contribution in realistic interior models, and to consider the contribution of inertial waves.
\end{abstract} 
\vspace{-1cm}
\vspace{-1cm}

\section{Introduction}\label{sec:intro}

The study of planetary interiors is an active and exciting area of research in which our understanding is rapidly developing, partly stimulated by recent observations from Juno and Cassini. Using observational evidence coupled with theoretical studies we can build on existing models. One such technique involves measuring orbital migration and spin evolution of star, planet, and moon systems. The evolution of these systems strongly depends on the tidal dissipation rates in the orbiting bodies, which are understood to be dependent on internal structure \citep[e.g.][]{Ogilvie2004,Mathis2013,Ogilvie2014}. Therefore, analysing long-term evolution of spins and orbits can provide constraints on the interior structures of giant planets.

Recent observations provide evidence for significant orbital migration of the moons of Jupiter and Saturn; their moons are migrating outwards at rates that require efficient tidal dissipation inside these planets \citep{Lainey2009,Lainey2012,Lainey2017,Lainey2020}. There are multiple mechanisms that could contribute to the efficient rates of tidal dissipation observed; in reality, the solution is likely to involve a combination of them. One possibility, and the mechanism explored in this paper, is that stable stratification or semi-convective layers within the giant planet allows for the excitation and subsequent dissipation of tidally-forced internal waves.
Other contributions to enhanced dissipation include the excitation and dissipation of inertial waves in convective zones \citep{Ogilvie2004,Wu2005a,Wu2005b,Favier2014}. Resonant locking of tidal gravito-inertial modes may occur \citep{Fuller2016}, where systems evolve so as to remain locked in particular resonances, thereby significantly enhancing dissipation in linear theory. Non-linear wave interactions within bodies, such as wave breaking \citep[e.g.][]{Barker2010} could also enhance the dissipation. It is thought that the rocky/icy inner cores of some planets could lead to additional visco-elastic sources of dissipation \citep[e.g.][]{Remus2012}. Finally, although more relevant for evolved binary stars, the non-wavelike (equilibrium) tide can interact with turbulent convection to enhance dissipation, though this mechanism is not thought to be effective in the solar system \citep[e.g.][]{Goldreich1977,Duguid2019,Duguid2020}.

Historically, when discussing the internal structure of giant planets, scientists refer to the three-layer model. This consists of a rocky/icy core underneath a convective envelope of metallic hydrogen and helium, surrounded by an outer molecular hydrogen and helium envelope \citep{Stevenson1982,Guillot2005,Fortney2010}. Although this is a reasonable starting point, the sizes of each region, and the exact nature of the transitions between them are uncertain, and these models cannot explain all observations. 

Recently, studies are departing from this standard model to explore other possible interior structures, in particular the inclusion of stable stratification or non-adiabatic profiles \citep[e.g.][]{Leconte2012,Vazan2016,Berardo2017,Lozovsky2017,Vazan2018,Debras2019}. Measurements of Jupiter’s gravity field (in particular from the Juno mission) suggest that the mass of Jupiter is distributed throughout the interior regions of the planet, existing as a large dilute core that extends some way out into the region previously thought to be convective \citep{Wahl2017,Helled2017,Debras2019}. Observations of density waves in Saturn’s rings due to resonances with global oscillations are consistent with stably stratified layers \citep{Marley1993,Hedman2013, Fuller2014,Hedman2018,Mankovich2021,Dewberry2021}. The frequencies observed suggest that strong stable stratification exists within Saturn's deep interior. It is possible that over time erosion or dissolution of giant planet cores extends the stratified region of the interior \citep{Guillot2004,Wilson2012,Moll2017}, or that a large core is created during the formation of these planets \citep{Pollack1996,Wahl2017}. Additionally, it is thought that temperature and pressure at the metallic / molecular boundary could be consistent with those required for helium rain, thus forming a stably stratified layer in this region \citep{Stevenson1977,Nettelmann2015}. 

We expect there to be an unstable entropy gradient throughout significant portions of giant planet interiors, which would compete with any stable composition gradients that form. As a result of these competing gradients the region could become unstable to double-diffusive convection, which in turn could lead to the formation of semi-convective layers (e.g.~\citealt{Wood2013,Garaud2018}). These layers consist of well-mixed convective regions, separated by thin stably stratified interfaces, such that the entropy profile appears as a staircase structure. Although these cannot be detected directly (unlike in the Arctic oceans on Earth, e.g.~\citealt{Ghaemsaidi2016,Shibley2017}), the formation of layers can give rise to different physics from the traditional adiabatic models and affect the behaviour and subsequent evolution of the system \citep[e.g.][]{Leconte2012,Vazan2016,Berardo2017,Vazan2018,Debras2019}.  It is possible that semi-convective layers inhibit heat transport, a key factor in planet evolution, and could be a contributing factor to the inflated radii of some hot Jupiters \citep{Chabrier2007} or the high luminosity observed in Saturn given its present age \citep{Leconte2013}.

In this paper we consider regions of stable stratification or semi-convective layers allowing internal gravity waves (g-modes) or interfacial modes, and study the consequences for tidal dissipation. In \citet{Pontin2020} we analysed the free modes and transmission of such waves in a global spherical model, and  \cite{Andre2019} considered the dissipation trends in an idealised Cartesian box model. Here we extend these works, adopting similar methods to \citet{Ogilvie2009}, to consider the dissipative properties of internal waves with realistic tidal forcing. We employ an idealised planetary model that allows considerable analytical understanding.

As in the previous studies we find a strongly frequency-dependent dissipation profile \citep[e.g.][]{Ogilvie2004,Ogilvie2009,Andre2019} that is strongly altered by the properties of the stratified regions \citep{Belyaev2015,Sutherland2016, Andre2017,Andre2019,Pontin2020}. We found that the properties of the stratification can vary both the location of resonances as well as the magnitude of the associated dissipation rate peaks. The consequences for a frequency-averaged or integrated dissipation quantity vary between parameter regimes but can lead to enhanced dissipation in some relevant cases. We find that although the results for semi-convective layers can depart from continuous uniform stratification when considering a small number of steps, the results quickly converge as step number increases. This initial study suggests that the presence of continuously stratified or semi-convective layers are valuable areas of investigation for tidal dissipation with potential to contribute significantly to tidal dissipation rates. In this study we have initially neglected rotation for simplicity and to enable further analytical development. However, we expect that rotation in planets is an important factor that would allow for inertial and gravito-inertial waves. These modes would also have an important contribution to the dissipation, but we leave this to a future paper.

The structure of this paper is as follows; in \S~\ref{sec:model} we outline our mathematical model and numerical methods. In \S~\ref{sec:overview} we discuss the overarching trends in the frequency-dependent tidal dissipation in some illustrative parameter regimes. In \S~\ref{sec:trav_wave} we consider how the low frequency regime behaves when waves are fully damped before forming a standing wave (``travelling wave"). In \S~\ref{sec:params} we examine how a frequency-averaged quantity for dissipation depends on the parameters of our model. In \S~\ref{sec:param_step}--\ref{sec:alt_g} we adapt the model to more specific scenarios such as semi-convective layers, helium rain, etc. In \S~\ref{sec4} we attempt to estimate the resulting astrophysical tidal timescales and quality factors. Finally in \S~\ref{sec:conc} we discuss our conclusions, and the implications of our results and highlight future areas of research.


\section{Model}\label{sec:model}

In this paper we aim to explore tidal dissipation due to internal waves in giant planets, examining how different forms of stable stratification might explain the enhanced dissipation observed. To do this we extend \citet{Andre2019}, who studied tidal dissipation in semi-convective layers in a Cartesian model, into a global spherical geometry, by extending the simplified model of \cite{Ogilvie2009}. 

In this initial study we adopt the idealised Boussinesq approximation \citep{Spiegel1960}, as this is the simplest computationally and allows for complementary analytical calculations. By doing this we are assuming that variations in density can be neglected except where they contribute to buoyancy and that the fluid is incompressible. This is appropriate when considering waves with shorter wavelengths than pressure or density scale heights, and with phase speeds that are slow relative to the sound speed. Although not strictly true for planetary-scale models this is likely to be a reasonable approximation for the dominant regime for tidally forced planetary waves. We also consider a uniform viscosity $\nu$ and thermal diffusivity $\kappa$. Analysis of the effects of rotation have not been included here, thereby first focussing on the non-rotating effects. The neglect of rotation is formally justified only if tidal forcing frequencies and buoyancy frequencies in any stable layers exceed those of the (mean) planetary rotational frequency $\Omega$. For Saturn, the typical values of the buoyancy frequency in the stable layer inferred by \cite{Mankovich2021} are much larger than $\Omega$, but the relevant tidal frequencies of interest for its moons are comparable in magnitude with $\Omega$. Hence, it is essential to build upon our study here to fully incorporate rotation before we can rigorously apply our model to tides in Jupiter and Saturn \citep[][]{Pontin2022}.

We build a global spherical model using spherical polar co-ordinates $(r,\theta,\phi)$, which are centred on the planet of mass $M$ and radius $r_0$. We neglect the distortion from spherical geometry caused by centrifugal effects and tidal forcing, and we use spherical harmonics to describe the angular structure of solutions. Neglecting tidal deformation is reasonable in giant planets due to the small tidal amplitudes. The neglect of rotational distortion, however, is more significant; although it would be possible to include this \citep[e.g.][]{Dewberry2022} we choose not to in this case as it would complicate analysis considerably, and is unlikely to significantly alter our conclusions. These approximations allow us to concentrate on studying the effects of gravity waves in isolation before adding additional effects into the model. 

\subsection{Governing equations}\label{sec:gov_eqn}
We consider the linearised form of the momentum equation, 
\begin{equation}\label{eq:mtm}
\frac{\partial \vb{u}}{\partial t} =-\frac{1}{\rho_0}\nablab p - b \vb{g} - \nablab \psi + \nu \nablab^2 \vb{u},
\end{equation}
where $\vb{u}$, $p$ are velocity and Eulerian pressure perturbations, $\rho_0$ the reference density and $\psi$ the tidal potential. We define the buoyancy variable $b$ to be $b =-\frac{\rho}{\rho_0}$, where $\rho$ is the density Eulerian perturbation. Note that this is not the standard definition for the buoyancy variable $b$ and does not have units of acceleration; instead it is a dimensionless quantity. This is done to allow us to define the radial dependence of the gravitational term separately, \hbox{$\vb{g}=-g(r) \boldsymbol{\hat{r}}$}, which allows for easy manipulation of the model to account for different gravity profiles. We focus predominantly on the gravity profile for a homogeneous body by considering $g=g_0 r$, where $g_0$ is the surface gravity with $r$ measured in units of planetary radius. We will discuss how results vary for a centrally condensed mass in \S~\ref{sec:alt_g}, where we consider $g=\frac{g_0}{r^2}$. We have neglected self-gravity to allow us to more easily explore our system analytically and because it is only likely to lead to a small linear effect on the quantitative results. 

We consider our system to be incompressible so conservation of mass gives 
\begin{equation}\label{eq:incompres}
\nablab \cdot \vb{u} = 0.
\end{equation}
The heat equation in terms of the buoyancy variable, $b$, becomes 
\begin{equation}\label{eq:heat}
\dpartial{b}{t} +\frac{u_r}{g}N^2 = \kappa \nablab^2 b,
\end{equation}
where $N^2$ is the radially dependent Brunt-V\"ais\"al\"a or buoyancy frequency, defined to be, 
\begin{equation}
N^2=g\left(\frac{1}{\Gamma_{1}}\frac{\mathrm{d} \ln p_0}{\mathrm{d}r}-\frac{\mathrm{d}\ln\rho_0}{\mathrm{d}r}\right),
\end{equation}
where $\Gamma_1=\left(\frac{\partial \ln p_0}{\partial \ln \rho_0}\right)_{\mathrm{ad}}$ is the first adiabatic exponent. This is equivalent to the system used in \citet{Pontin2020}, except for the addition of dissipative terms and realistic tidal forcing. As in that study, we approximate the buoyancy frequency to represent a density gradient as
\begin{equation}
N^2\approx-\frac{g}{\rho_0}\frac{\mathrm{d}\rho_0}{\mathrm{d}r}.
\end{equation}

We introduce tidal forcing by considering the dominant component of the tidal potential \citep{Ogilvie2014},
\begin{equation}\label{eq:forcing}
\psi = \psi_0~r^2 Y^2_2(\theta,\phi) e^{-i \omega t},
\end{equation}
where $\psi_0 =\sqrt{\frac{6 \pi}{5}}\frac{M_2}{M}\omega_d^2 \big(\frac{r_0}{a} \big)^3$. We take the $l=m=2$ dimensional tidal amplitude for a circular orbit as defined in \cite{Ogilvie2014}, where $M$ is the mass of the planet, $M_2$ is the companion mass, $a$ is the orbital semi-major axis, and the dynamical frequency $\omega_d=\sqrt{GM/r_0^3}$, where $G$ is the gravitational constant. The forcing frequency is $\omega=2(\Omega_o-\Omega_s)$, where $\Omega_o$ is the orbital frequency of the satellite and $\Omega_s$ the spin frequency of the planet (note here $\Omega_s=0$). This is the most relevant one for a circularly aligned orbit of a non-synchronously orbiting moon. Note that we solve for the entire linear tidal response to $\psi$ directly; we do not need to split up the tide into an equilibrium and a dynamical tide, though only the dynamical/wavelike response is typically important for the dissipation in our model.

We expand perturbations using spherical harmonics with a harmonic time-dependence such that, 
\begin{equation}\label{eq:u_r}
u_r(r,\theta,\phi,t)= \tilde{u}_r^l(r) Y^{m}_{l}(\theta,\phi)e^{-i\omega t}, 
\end{equation}
\begin{equation}\label{eq:u_theta}
u_{\theta}(r,\theta,\phi,t)= r \tilde{u}_b^l(r) \dpartial{}{\theta} Y^{m}_{l}(\theta,\phi)e^{-i\omega t},
\end{equation}
\begin{equation}\label{eq:u_phi}
u_{\phi}(r,\theta,\phi,t)= r \frac{\tilde{u}_b^l(r)}{\sin \theta} \dpartial{}{\phi} Y^{m}_{l}(\theta,\phi)e^{-i\omega t},
\end{equation}
\begin{equation}
p(r,\theta,\phi,t)= \tilde{p}^l(r) Y^{m}_{l}(\theta,\phi)e^{-i\omega t},
\end{equation}
\begin{equation}
b(r,\theta,\phi,t)= \tilde{b}^l(r) Y^{m}_{l}(\theta,\phi)e^{-i\omega t},
\end{equation}
where spherical harmonics are normalized such that, 
\begin{equation*}
\int_0^{2\pi} \int_0^\pi  [Y_{l'}^{m'} (\theta,\phi)]^* Y_l^m(\theta,\phi)  \sin^2\theta\, \mathrm{d} \theta\, \mathrm{d} \phi=\delta_{l,l'}\, \delta_{m,m'}.
\end{equation*}

The resulting equations are (where again tildes have been dropped for clearer notation) 
\begin{multline}
(-i \omega) u_r^l \\ = -\frac{1}{\rho_0} \dtotal{p^l}{r} + g b^l - \dtotal{\psi^l}{r}\delta_{l,2} -\nu \frac{l(l+1)}{r^2} \Big[u_r^l - \dtotal{}{r}(r^2 u_b^l)\Big],
\end{multline}
\begin{multline}
(-i \omega) r^2 u_b^l \\ = -\frac{p^l}{\rho_0}-\psi^l \delta_{l,2} + \nu\, \Bigg[ \frac{2 u_r^l}{r}+\frac{1}{r^2}\dtotal{}{r} \bigg( r^4 \dtotal{u_b^l}{r} \bigg) -(l-1)(l+2) u_b^l \Bigg],
\end{multline}
\begin{equation}
\frac{1}{r^2}\dtotal{}{r}(r^2 u_r^l)-l(l+1)u_b^l=0,
\end{equation}
\begin{equation}
(-i \omega)b^l+N^2\frac{u_r^l}{g} = \kappa\, \Bigg[ \frac{1}{r^2}\dtotal{}{r}\bigg(r^2 \dtotal{b^l}{r} \bigg) - \frac{l(l+1)}{r^2}b^l \Bigg],
\end{equation}
consistent with those in \cite{Ogilvie2009}. 
The equations for each $m$ are uncoupled due to the axisymmetric basic state, and due to the neglect of the Coriolis term components with different angular wavenumbers $l$ are also uncoupled. Hence, we solve only for $m=2$ and $l=2$. 

We non-dimensionalise our system using units of length in terms of the planetary radius $r_0$, the reference density $\rho_0$ for density, and our unit of time $\omega_d^{-1}$. Therefore, we introduce the following dimensionless parameters,  $r=r_0\hat{r}$, $u_r=r_0 \omega_d \hat{u}_r$, $u_b=\omega_d \hat{u}_b$, $u_c=\omega_d \hat{u}_c$, $\psi=\psi_0\hat{\psi}$, $b=\hat{b}$, $g=g_0\hat{g}$, $p=\rho_0 r_0 g\hat{p}$, $\omega=\sqrt{\frac{g_0}{r_0}}\hat{\omega}$, where $g_0=\omega_d^2 r_0$ is the surface gravity. We then drop the hats to simplify notation, giving us, 
\begin{multline}\label{eq:ur_nondim}
(-i \omega) u_r^l = -\frac{1}{\rho_0} \dtotal{p^l}{r} + gb - \frac{\psi_0}{r_0 g_0} \dtotal{ \psi^l}{r}\delta_{l,2} \\ -\frac{\nu}{\sqrt{r_0^3 g_0}} \frac{l(l+1)}{r^2}\Bigg[u_r^l - \dtotal{}{r}(r^2 u_b^l)\Bigg],
\end{multline}
\begin{multline}
(-i \omega) r^2 u_b^l = -\frac{p^l}{\rho_0} -\frac{\psi_0}{r_0 g_0} \psi^l\delta_{l,2} \\ + \frac{\nu}{\sqrt{r_0^3 g_0}} \Bigg[ \frac{2 u_r^l}{r}+\frac{1}{r^2}\dtotal{}{r} \bigg( r^4 \dtotal{u_b^l}{r} \bigg) -(l-1)(l+2) u_b^l \Bigg],
\end{multline}
\begin{equation}
\frac{1}{r^2}\dtotal{}{r}(r^2 u_r^l)-l(l+1)u_b^l=0,
\end{equation}
\begin{equation}\label{eq:b_nondim}
(-i \omega)b^l+\frac{N^2r_0}{g_0}\frac{u_r^l}{g} =\frac{\kappa}{\sqrt{r_0^3 g_0}} \Bigg[ \frac{1}{r^2}\dtotal{}{r}\bigg(r^2 \dtotal{b^l}{r} \bigg) - \frac{l(l+1)}{r^2}b^l \Bigg].
\end{equation}
We highlight that we have three key dimensionless parameters for each frequency $\omega$,
\begin{equation*}
\frac{N^2}{\omega_d^2},~~~
\frac{\nu}{r_0^2 \omega_d}~~~\textrm{and}~~~\frac{\kappa}{r_0^2 \omega_d}~~~\bigg(\textrm{or alternatively } Pr=\frac{\nu}{\kappa}\bigg),
\end{equation*}
which are varied in later analysis, and referred to simply as $N^2$ (which can be a function of radius), $\nu$ and $Pr$ as we used units of $r_0=1$ and $\omega_d=1$. To aid understanding we include only dimensional equations in later sections and figures are plotted using these units. 

\subsection{Boundary Conditions}\label{sec:boundary}

In our numerical calculations we enforce boundary conditions at either end of the domain, which extends from a non-zero inner core boundary $\alpha r_0$ to the planet's radius $r_0$. We assume there is a rigid inner core that remains spherical, thereby neglecting any deformation caused by rotation or tidal forcing.  Therefore, we impose no normal flow at the inner core, 
\begin{equation}
u_r^l= 0 \quad \mathrm{at}  \quad  r=\alpha r_0. 
\end{equation} 
At the tidally-perturbed outer boundary we consider it to be a free surface on which the normal stresses vanish.  Therefore, we consider the perturbations to the normal stress to include the Lagrangian pressure perturbation $\Delta p$, and the normal viscous stress, i.e.
\begin{equation}
\Delta p - 2 \nu e_{rr} = (\tilde{p}+\vb{\xi} \cdot \nabla p_o) - 2 \nu e_{rr} = 0,
\end{equation} 
where $\tilde{p}$ is the Eulerian perturbation for this equation only and $p_0(r)$ is the hydrostatic pressure profile. This can be evaluated to give the boundary condition, 
\begin{equation}
W^l - \frac{g_0}{(-i \omega)} u_r^l - 2 \nu \frac{\mathrm{d} u_r^l}{\mathrm{d} r} = \psi_{0} \delta_{l2} \quad \mathrm{at}   \quad  r=r_0,
\end{equation}
where $g_0$ is the surface gravity, which can be related to the dynamical frequency $g_0=-\omega_d^2 r_0$, and $W^l=\frac{p^l}{\rho_0}+\psi$, neglecting self-gravity. 

At both boundaries, we consider stress-free conditions (no tangential stress), which is consistent with a free surface. Realistically we expect the inner core would be closer to non-slip but it is numerically convenient to consider stress-free conditions without it having any significant effect on the results. This means that
\begin{equation}
\frac{\mathrm{d} u_b^l}{\mathrm{d}r}+\frac{u_r^l}{r^2}=0 \quad \mathrm{at}  \quad  r=\alpha r_0\quad \mathrm{and} \quad  r=r_0.
\end{equation}
We considered different boundary conditions on the buoyancy variable; however, we found that there was no significant effect on the results as long as the condition chosen did not contradict the governing equations. For the results shown in this study, we have considered the inner core to be fixed entropy, such that there is no perturbation to the quantity $b$, i.e.
\begin{equation}
b^l=0 \quad \mathrm{at} \quad r=\alpha r_0, 
\end{equation}
and no perturbation to the buoyancy flux through the surface, i.e.
\begin{equation}
\frac{\partial b^l}{\partial r}=0  \quad \mathrm{at} \quad r=r_0. 
\end{equation}

\subsection{Energy Balance} \label{sec:energy}

As we aim to understand the dissipation of tidally forced internal waves, we turn our attention to the energetic quantities and their balances. Considering the standard definition for work, the mean rate of energy injection by the tidal forcing is defined to be 
\begin{equation}
I=\int_V \rho_0 \big(\vb{u}\cdot \vb{F}\big) \mathrm{d}V, 
\end{equation}
where $\vb{F}=-\nabla \psi$ is the tidal acceleration. 

Therefore, by taking the scalar product of equation \ref{eq:mtm} with $\rho_0 \vb{u}$, using equation \ref{eq:heat} and integrating over the volume, the following energy balance is found,
\begin{equation}\label{eq:energy_balance}
I = \frac{\mathrm{d} E_K}{\mathrm{d} t}+\frac{\mathrm{d} E_{PE}}{\mathrm{d} t} +\frac{1}{V} \oint_S p \vb{u} \cdot \mathrm{d}\vb{S}+D_{ther}+D_{visc}.
\end{equation}
$E_k$ and $E_{PE}$ are the kinetic and potential energy of the system respectively, where the rate of change of kinetic energy is,  
\begin{equation}\label{eq:KE}
\frac{\mathrm{d} E_K}{\mathrm{d} t} = \int_V \frac{\rho_0}{2}  \frac{\partial|\vb{u}|^2}{\partial t}  \mathrm{d}V,
\end{equation}
and the rate of change of potential energy,
\begin{equation}\label{eq:PE}
\frac{\mathrm{d} E_{PE}}{\mathrm{d} t} = \int_V \frac{g^2}{2 N^2} \frac{\partial b^2}{\partial t} \mathrm{d} V,
\end{equation}
except when $N=0$, in which case $E_{PE}=0$ and $\dtotal{E_{PE}}{t}=0$. The volume integrated viscous dissipation rate is given by,
\begin{equation}\label{eq:Dvisc}
D_{visc}=-\int_V \rho_0 \nu \vb{u} \cdot \nabla^2 \vb{u} \mathrm{d} V,
\end{equation}
and volume integrated thermal dissipation rate is written, 
\begin{equation}\label{eq:Dther}
D_{ther}=-\int_V \rho_0 \kappa \frac{g^2}{N^2} b \nabla^2 b \mathrm{d} V,
\end{equation}
again as for potential energy when $N=0$, $D_{ther}=0$.

\begin{figure*}[t]\centering
       \subfigure{
		\begin{tikzpicture}[scale=2]
\tikzstyle{every node}=[font=\small]
		\draw[->] (0,0) -- (1.2,0) coordinate (x axis) node[right]{\textcolor{red}{$N$} \textcolor{blue}{$s$}};
		\draw[->] (0,0) -- (0,2.4) coordinate (y axis) node[above]{$r$};
		\foreach \y/\ytext in {0/0, 0.3/ \alpha r_0, 0.9/ \beta r_0, 2.1/r_0} 
		\draw (1pt,\y cm) -- (-1pt,\y cm) node[anchor=east,fill=white] {$\ytext$};
		\draw[thick,blue] (0.4,0.9) -- (0.95,0.3);
		\draw[thick,blue] (0.4,0.9) -- (0.4,2.1);
		\draw[ultra thick,red,dashed] (0.7,0.3) -- (0.7,0.9);
		\draw[ultra thick,red,dashed] (0.01,0.9) -- (0.7,0.9);
		\draw[ultra thick,red,dashed] (0.01,0.9) -- (0.01,2.1);
		\node at (1.75,0.15) {Inner Core};		
		\node at (1.75,0.6) {Outer Core};
		\node at (1.75,1.5) {Convective Region};
		\node at (1.7,2.1) {Planet Radius};
		\draw[->] (2.5,0) -- (3.7,0) coordinate (x axis) node[right]{\textcolor{red}{$N$} \textcolor{blue}{$s$}};
		\draw[->] (2.5,0) -- (2.5,2.4) coordinate (y axis) node[above]{$r$};
		\foreach \y/\ytext in {0/0, 0.3/ \alpha r_0, 0.9/ \beta r_0, 2.1/r_0} 
		\draw (2.49,\y cm) -- (2.49,\y cm) node[anchor=east,fill=white] {$\ytext$};
		\draw[thick,blue] (3.5,0.5) -- (3.5,0.3);
		\draw[thick,blue] (3.3,0.5) -- (3.5,0.5);
		\draw[thick,blue] (3.3,0.7) -- (3.3,0.5);		
		\draw[thick,blue] (3.1,0.7) -- (3.3,0.7);
		\draw[thick,blue] (3.1,0.9) -- (3.1,0.7);
		\draw[thick,blue] (2.9,0.9) -- (3.1,0.9);
		\draw[thick,blue] (2.9,0.9) -- (2.9,2.1);
		\draw[ultra thick,red,dashed] (2.51,0.46) arc (-90:90:0.5cm and 0.04cm);
		\draw[ultra thick,red,dashed] (2.51,0.66) arc (-90:90:0.5cm and 0.04cm);
		\draw[ultra thick,red,dashed] (2.51,0.86) arc (-90:90:0.5cm and 0.04cm);
		\draw[ultra thick,red,dashed] (2.51,0.3) -- (2.51,0.46);
		\draw[ultra thick,red,dashed] (2.51,0.54) -- (2.51,0.66);
		\draw[ultra thick,red,dashed] (2.51,0.74) -- (2.51,0.86);
		\draw[ultra thick,red,dashed] (2.51,0.94) -- (2.51,2.1);
		\end{tikzpicture}}
	\subfigure{\includegraphics[width=0.21\textwidth, trim={3.5cm 0 3.5cm 0},clip]{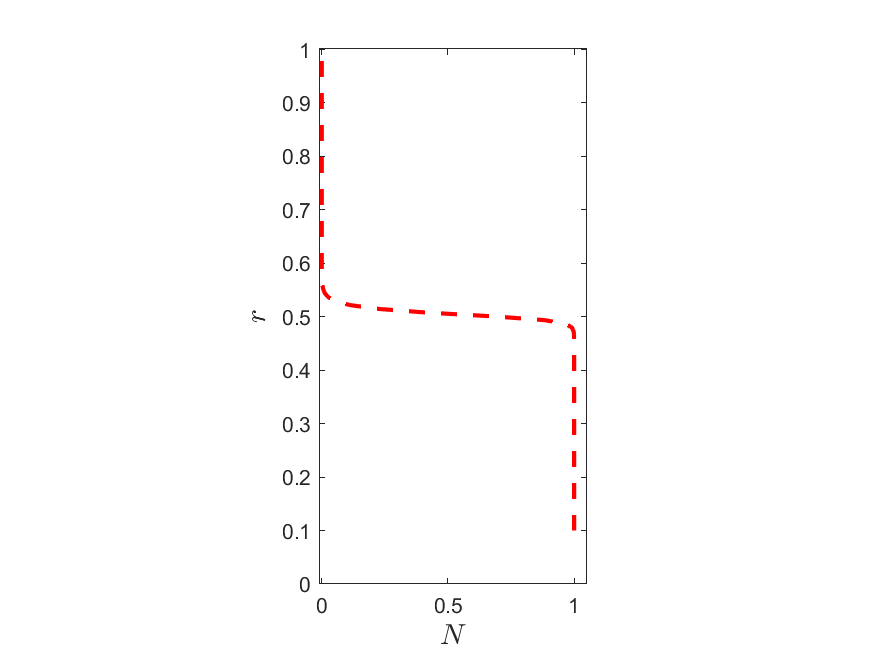}}
	\subfigure{\includegraphics[width=0.21\textwidth, trim={3.5cm 0 3.5cm 0},clip]{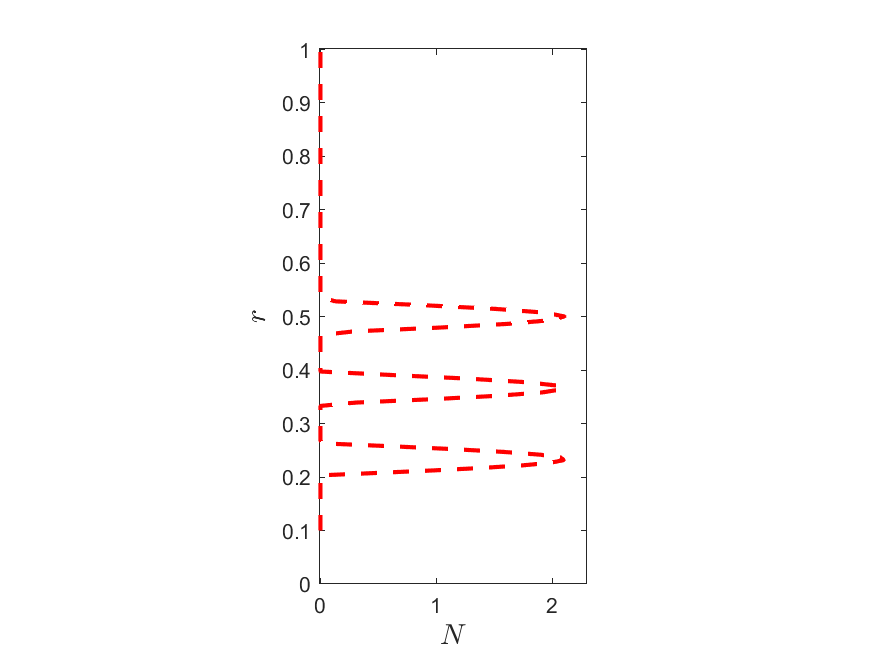}}
	\caption{Left panels: Illustrative examples of the entropy profile and  Brunt-V\"ais\"al\"a frequency ($N^2$) for both a continuously stratified layer and semi-convective layers.  In both cases there is a solid inner core that extends to $\alpha r_0$, outer core containing stable stratification extending to $\beta r_0$, and planetary radius $r_0$. Right panels: examples of the numerically smooth profiles used.}\label{fig:profile_ex}
\end{figure*}
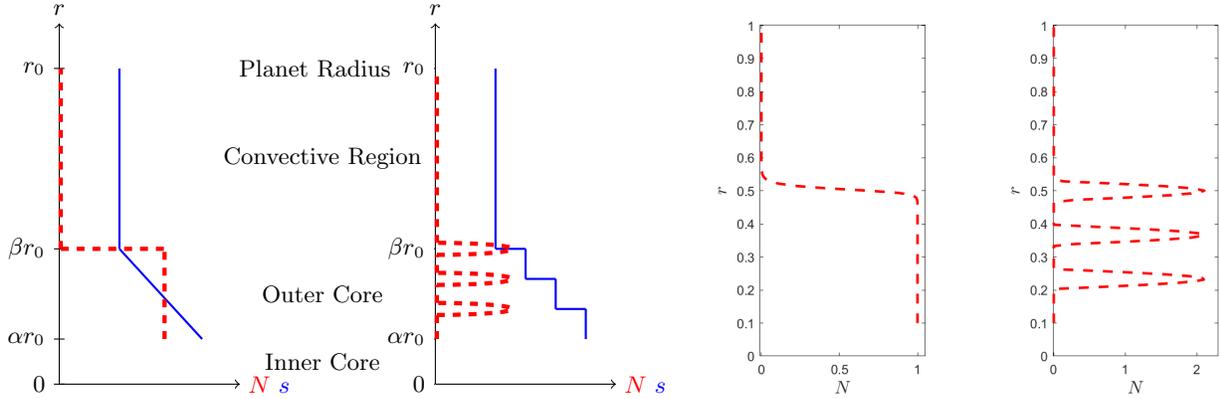

In Appendix \ref{app:dvisc}, we show how the viscous dissipation can be divided into two separate components. These consist of the viscous dissipation within the bulk of the fluid and the normal viscous stresses. The viscous dissipation within the bulk can be written \citep{Ogilvie2009},
\begin{multline}\label{eq:bulk_visc}
D_{interior}= \frac{\rho_0 \nu}{2} \sum_l l(l+1)\Bigg( \bigg|\frac{u_r^{l}}{r}+r\dtotal{u_b^l}{r}\bigg|^2 + 3 \bigg|\dtotal{u_r^l}{r}\bigg|^2 \\ 
+(l-1)l(l+1)(l+2)|u_b^l|^2 \Bigg).
\end{multline}
Note in this case we are only considering $l=2$. From now on, when considering viscous dissipation we refer to the quantity defined in equation \ref{eq:Dvisc}, but with the parameters we consider differences between these quantities are negligible. 

Using the divergence theorem, the mean energy injection rate from the tidal forcing can be written in the form,
\begin{equation}
I= \int_V \rho_0 \vb{u}\cdot (- \nabla \psi) \mathrm{d}V = -\oint_S \rho_0 \psi \vb{u} \cdot \mathrm{d}\vb{S}.
\end{equation}

In a steady state, averaged over the forcing period of $\frac{2 \pi}{\omega}$, the kinetic and potential energy terms (equations \ref{eq:KE} and \ref{eq:PE}) will be identically zero. Therefore, the injection rate $I$ will balance the total viscous and thermal dissipation terms, $D_{visc}$ and $D_{ther}$. It can be shown numerically that an additional balance between the dissipation within the bulk of the fluid defined in equation~\ref{eq:bulk_visc}, and the pressure integral in equation \ref{eq:energy_balance} also exists (see also Appendix A). In this study we will analyse how the dissipation rates $D_{visc}$ and $D_{ther}$ depend on other properties of the system. For some analysis in this paper $D_{ther}$ has been calculated by considering this balance, as it is less numerically demanding to reach a converged result\footnote{The solutions of our tidally-forced linear boundary value problem have been checked to be very accurately converged numerically with our adopted resolutions by comparing results obtained with different numbers of radial grid points. The issue we refer to here is purely in the post-processing analysis to obtain $D_{ther}$ for cases with spatially variable $N^2$ profiles, and results from the division by $N^2$ (where a threshold value must be imposed) and in the radial derivatives of $b$ involved when using $N^2(r)$ profiles that are not infinitely differentiable.}.

\subsection{Density structure}

We consider regions in a giant planet where stable layers have formed, or which could have evolved into semi-convective layers due to double-diffusive convection. As discussed in \S~\ref{sec:intro} there are two regions to consider, one close to the core, or alternatively  near the H/He molecular to metallic transition radius.

To incorporate both continuous stable stratification and semi-convective layers in our model, as well as considering multiple locations for these layers, we define the buoyancy frequency in several ways. First, we consider a density profile representative of a continuous stably stratified region, with a constant $N$ which sits above the solid core extending a defined distance into the planetary envelope. This is shown on the left hand side of Figure~\ref{fig:profile_ex}. This is described by a step function in the buoyancy profile $N^2(r)$ that is non-zero from the inner core boundary $\alpha r_0$ to the outer core boundary $\beta r_0$. For our numerical calculations, when the stable layer does not extend to the planetary surface ($\beta \neq 1$), we consider a smoothly varying buoyancy profile, 
\begin{equation}\label{eq:N2_layer}
N^2(r)=\frac{\bar{N}^2}{2} \Big( \tanh \big( \Delta(\beta r_0-r) \big)+1 \Big).
\end{equation}
Unless specified otherwise we set $\Delta=100$. For numerical reasons, the buoyancy term is set to be identically zero away from the step, when $N^2(r) < \frac{\bar{N}^2}{10^7}$. For cases where $\beta=1$ the buoyancy profile is constant throughout the domain, i.e.,
\begin{equation}\label{eq:N2_layer_beta1}
N^2(r)=\bar{N}^2.
\end{equation}

The second density structure explored represents a semi-convective structure with $n_{max}$ steps within the stratified layer. We want to consider a series of $\delta$-functions in the buoyancy variable to give a staircase-like density profile, shown on the right hand side of Figure \ref{fig:profile_ex}. To do this numerically, we consider finitely thin and smoothly varying interfaces by taking
\begin{equation}\label{eq:N2_steps}
N^2(r)=
\begin{cases}
\frac{N_0^2}{2} \bigg(1+\cos\Big(2 \pi \frac{r-r_n}{\delta r}\Big)\bigg)  \quad & |r-r_n| < \frac{\delta r}{2},\\
0 \quad & \text{otherwise},
\end{cases}
\end{equation}
for $1<n<n_{max}$, where $r_n=\alpha r_0+nd$, $d=\frac{(\beta-\alpha)r_0}{n_{max}}$, and $\epsilon=\frac{\delta r}{d}$. $N_0^2$ is set to a value which gives a mean stratification for the staircase equivalent to a region with constant stratification $\bar{N}^2$,
\begin{equation}
N_0^2 = \bar{N}^2 \frac{(\beta - \alpha) r_0}{n_{max}~\delta r}.
\end{equation}
This allows for comparison between a stratified layer and semi-convective region. Note that if $\beta=1$, the profile is adjusted such that the final step isn't included at the planetary radius ($0<n<n_{max}-1$), and $N_0$ is adjusted to maintain the mean stratification. 

Finally, the formulation in equation \ref{eq:N2_steps} can also be used to represent a stratified layer at the metallic/molecular transition zone, where instead of multiple steps we consider one wide step in the transition region,  
\begin{equation}
N^2(r)=
\begin{cases}
\frac{N_0^2}{2} \bigg(1+\cos\Big(2 \pi \frac{r-\beta}{\delta r} \Big)\bigg)   \quad &  |r-\beta| < \frac{\delta r}{2},\\
0 \quad & \text{otherwise},
\end{cases}
\end{equation}
and 
\begin{equation}
N_0^2 = \bar{N}^2 \frac{(r_0 - \alpha) r_0}{\delta r}.
\end{equation}
This single step can be used as an isolated stable layer embedded within a convective medium. The choice of $N_0$ to compare with a stratified layer across the entire domain is somewhat arbitrary as it is not insightful here to make comparisons between these quantities. 

In our model we have adopted a single density variable $b$ with a single ``thermal diffusivity" $\kappa$ to model our density/entropy stratification for simplicity. If compositional gradients are important (or even dominant) then in reality the density should consist of two distinct components (thermal and compositional) which diffuse at different rates. This is indeed what is required for double-diffusive convection to occur in giant planet interiors and to produce a layered/staircase density structure \citep[e.g.][]{Garaud2018}. In our model this could be accounted for by having two different buoyancy variables with different associated diffusivities \citep[e.g.][who estimate the compositional diffusivity to be comparable with $\nu\approx 10^{-2}\kappa$ in Jupiter]{Moll2017}. It would be worthwhile to explore such double-diffusive effects in future work to determine their effects on the dissipative tidal response.

\subsection{Frequency averaged dissipation}

Although the response to tidal forcing is known to be strongly dependent on the forcing frequency \citep[e.g.][]{Ogilvie2004, Fuller2016, Andre2019}, to fully explore the system's dependences on the parameters and functional form of the stratification profile, it is helpful to define a quantity that gives a quantitative measure of the dissipation that can be compared as the parameters are varied. We consider a frequency-averaged dissipation as a measure of overall dissipation \citep{Ogilvie2014},
\begin{equation}\label{eq:freq_avg}
\bar{D} = \int^{\omega_{max}}_{\omega_{min}} \frac{D(\omega)}{\omega}~\mathrm{d} \omega.
\end{equation} 
This gives more emphasis to the lower frequencies which are expected to be more significant when considering a tidally forced system. We take a small non-zero limit for the lower bound of the integral numerically, found by checking for convergence of the results. Unless otherwise stated, $\omega_{max}$ is taken to be $\bar{N}$. This allows analysis of the low frequency regime without results being quite as dominated by the surface gravity (f-mode) behaviour. Removing the surface gravity mode and this limit and the $\frac{1}{\omega}$ weighting is appropriate to study planet-satellite systems where we expect the frequencies of tidal forcing to be small compared to the dynamical frequency of the body. Other weighting factors could be used for the frequency-averaged dissipation instead of $\frac{1}{\omega}$, such as the $\frac{1}{\omega^2}$ weighting used in \cite{Ogilvie2013}. However, we found it does not alter our overall conclusions so choose to present results for this weighting only. 

\subsection{Comparison to a simple harmonic oscillator}\label{sec:SHO}

Parallels exist between the response of a body to a tidal potential and that of a forced, damped, simple harmonic oscillator. For this reason, a simple harmonic oscillator is often used as an analogy for a planet-moon system and is helpful for understanding some of our later results. Considering a forced, damped, simple harmonic oscillator for a quantity $a(t)$ with a single resonant frequency $\omega_0$, we can write, 
\begin{equation}
\ddtotal{a}{t}+\omega_0^2 a = \hat{F} \cos \omega t - \epsilon \omega_0 \dtotal{a}{t},
\end{equation}
where $\hat{F} \cos \omega t$ is the forcing and $\epsilon \omega_0 \dtotal{a}{t}$ is the damping. 
By considering a solution of the form $\operatorname{Re}[\hat{a}~e^{-i\omega t}]$, dissipation can be shown to be,  
\begin{equation}
D = \bigg \langle \epsilon \Big( \dtotal{a}{t} \Big)^2 \omega_0 \bigg \rangle \propto \frac{\epsilon \omega_0 \omega^2}{(\omega_0^2 - \omega^2)^2 + \epsilon^2 \omega_0^2 \omega^2}.
\end{equation}
This exhibits a resonance at $\omega = \pm \omega_0$, and if we consider the dissipation at these frequencies, we find 
\begin{equation}
D_{max} \propto \frac{1}{\epsilon \omega_0},
\end{equation}
showing there is a clear inverse relationship between the damping rate $\epsilon$ and the peak dissipation. 

We define the edge of a peak to be the frequency at which the dissipation has decreased to half the maximum peak height. The frequencies either side of $\pm \omega_0$ where this value is reached are,
\begin{equation}
\omega_{+,\frac{1}{2}}=\frac{\omega_0}{2} \bigg(\pm \epsilon + \sqrt{\epsilon^2 +4} \bigg),
\end{equation}
and similarly for $-\omega_0$. Therefore the half-width of the peak $\Delta$ when $\epsilon \ll 1$ is approximately  
\begin{equation}
\Delta = \epsilon \omega_0,
\end{equation}
showing there is a linear relationship between damping rate and peak width. We refer to these two relationships when discussing our findings in \S~\ref{sec:results}. 

\subsection{Eigenvalue problem} \label{sec:eigs}

It is informative to see how the dissipation patterns correspond with the expected free modes of the system. Although we gained some understanding of the possible free modes for the inviscid case in \citet{Pontin2020}, we can also examine these by considering the numerical eigenvalues of the dissipative system. 

To do this we consider the unforced case where \hbox{$\psi_0=0$,} which becomes the following generalised linear eigenvalue problem with eigenvalue $(-i\omega)$, described by equations \ref{eq:ur_nondim} to \ref{eq:b_nondim} with \hbox{$\psi_0=0$} and with the same boundary conditions. This is then solved using an inbuilt MATLAB linear algebra routine, to give the eigenvalues ($-i\omega$), and the corresponding eigenvectors for $\vb{u}$, $p$, $b$. We can then consider the $\operatorname{Re}[\omega]$ and $\operatorname{Im}[\omega]$ parts separately as they are the frequency of the mode, and the associated damping rate, respectively. We use the iterative “eigs” method to scan the relevant frequency range to reduce computing requirements.

\begin{figure}
\centering	
\subfigure[$\alpha=0.1$,  $\beta=1.0$,  $\bar{N}^2=1$,  $\nu=\kappa=10^{-6}$]{\includegraphics[width=0.46\textwidth]{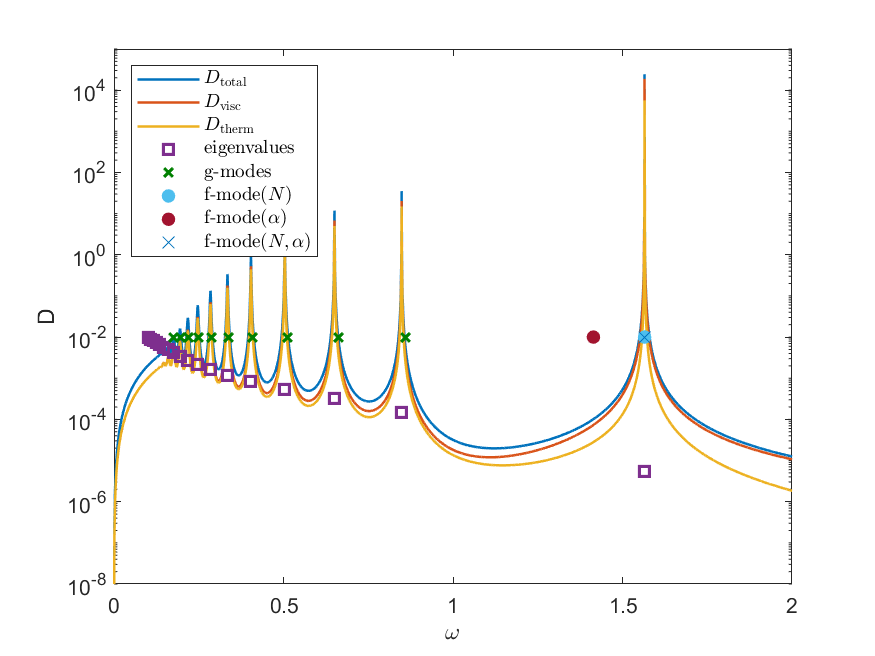}\label{fig:diss_over_a}}
\subfigure[$\alpha=0.1$,  $\beta=0.5$,  $\bar{N}^2=1$,  $\nu=\kappa=10^{-6}$]{\includegraphics[width=0.46\textwidth]{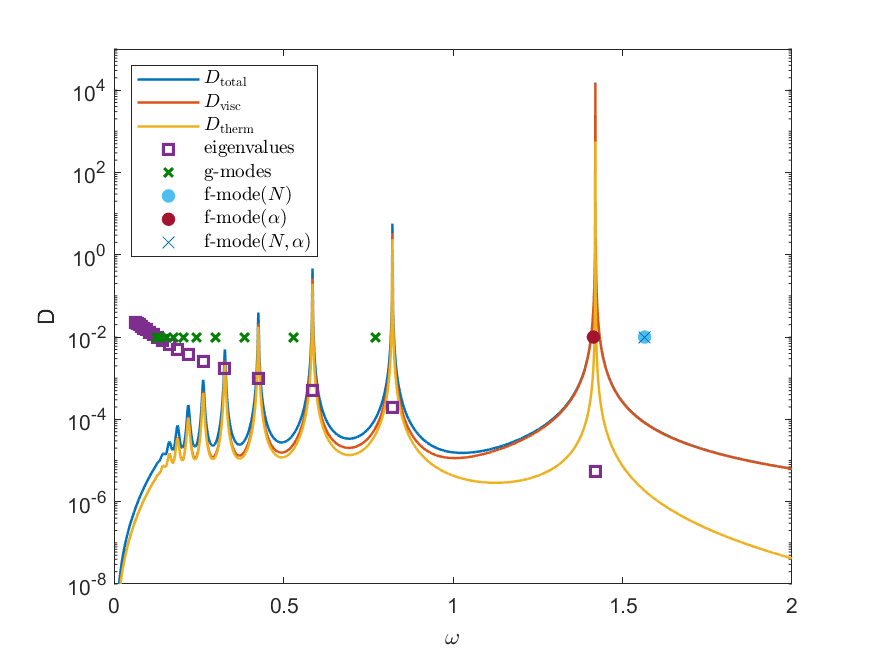}\label{fig:diss_over_b}}
\subfigure[$\alpha=0.1$,  $\beta=0.5$,  $\bar{N}^2=1$,  steps$=3$,  $\delta r=0.06$, $\nu=\kappa=10^{-6}$]{\includegraphics[width=0.46\textwidth]{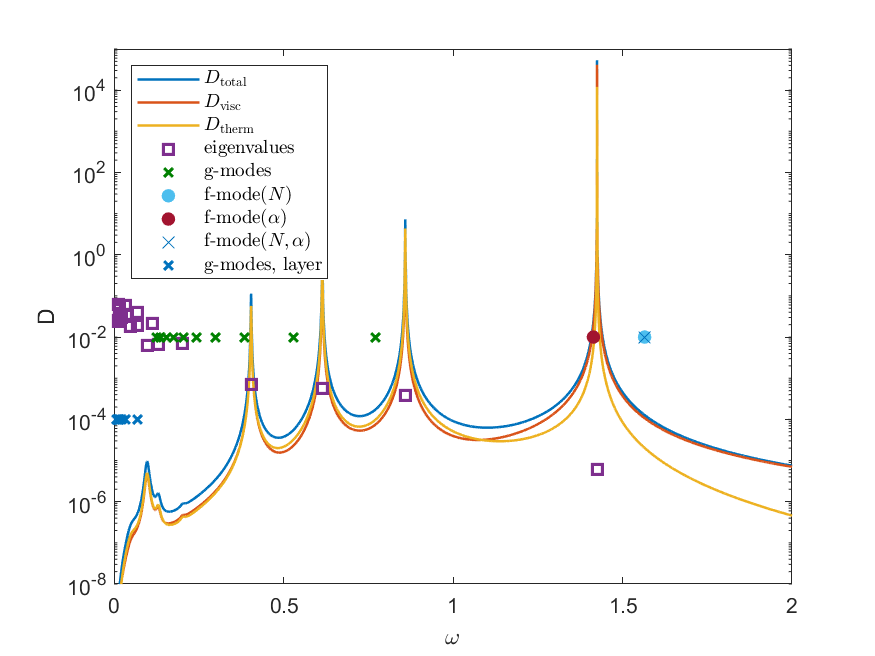}\label{fig:diss_over_c}}
\caption{Illustrative examples of the dissipation as a function of frequency together with eigenvalue solutions and analytical calculations for two cases considering uniform stratification (panels \protect\subref{fig:diss_over_a} and \protect\subref{fig:diss_over_b}), and one considering semi-convective layers (panel \protect\subref{fig:diss_over_c}). The $y$-axis values for analytical results are arbitrary and those for the eigenvalues are the associated damping rates.} \label{fig:diss_over}
\end{figure}

\subsection{Numerical method}

We solve the system of ordinary differential equations in radius, equations~\ref{eq:ur_nondim} to~ \ref{eq:b_nondim} for each $l$ using a Chebyshev collocation method, where the ordinary differential equations in $r$ are converted into a linear system of equations on a Chebyshev grid \citep{Trefethen2000,Boyd2001}.  

The angular dependence is handled by employing spherical harmonics, and here we focus on $l=2$, and note that different $l$'s are uncoupled in our linear calculations without rotation. We consider points in radius as a set of $(n_{cheb}+1)$ Gauss-Lobatto-Chebyshev points. The value of $n_{cheb}$ varies with choice of density structure and parameters, but we typically take $n_{cheb}=100$ to $400$. These are an appropriate choice of basis for many non-periodic problems and have been shown to have good convergence properties. In particular, spectral methods such as this converge exponentially fast with resolution $n_{cheb}$ for smooth solutions \citep{Boyd2001}. 

The resulting linear algebra problem is solved using the inbuilt MATLAB routine “mldivide”, where matrices are stored in sparse form to reduce the numerical memory requirements. The solutions for $\vb{u}$, $p$ and $b$ can then be used in equations \ref{eq:Dvisc} and \ref{eq:Dther}.


\section{Results}\label{sec:results}

Having established the framework for our numerical work we now consider the numerical and analytical results for the case of a non-rotating body. This will allow us to gain an initial understanding of the system with reduced numerical cost, and complementary analytical calculations. We will discuss the basic properties of our system before considering how the dissipative properties depend on the model's key parameters, and discuss the implications for astrophysical tidal evolution. 

\subsection{Outline of the key features}\label{sec:overview}

First, we summarise the overarching trends that we observed when evaluating the frequency dependence of the viscous, thermal and total dissipation, defined by equations~\ref {eq:Dvisc}~and~\ref{eq:Dther}. We know from previous studies, both with and without rotation, that the magnitude of the dissipation has a strong dependence on the forcing frequency \citep[e.g.][]{Ogilvie2004,Ogilvie2009,Andre2019} and indeed we find this again here. 

Illustrative examples for three different profiles are shown in Figure~\ref{fig:diss_over}, where we can see a strong frequency dependence. This shows viscous ($D_{visc}$), thermal ($D_{therm}$), and total ($D_{visc}+D_{therm}$) dissipation rates as a function of the forcing frequency ($\omega$). The eigenvalues for the frequencies of the free modes have been found using the methods described in \S~\ref{sec:eigs}. These are shown by the purple squares and we plot the frequency of the mode ($\operatorname{Re}[\omega]$) on the $x$-axis and the damping rate ($\operatorname{Im}[\omega]$) on the $y$-axis. The green crosses show the analytical calculation for the frequencies of the g-modes, as derived in Appendix~\ref{app:g-mode}. Circles show the analytical calculation for the frequencies of the f-mode, as derived in Appendix~\ref{app:f-mode}. For all analytical results the $y$-axis values are arbitrary. Dissipation $D$ is measured in units of $\rho_0 r_0^5 \omega_d^3$, and $\omega$ is shown in units of $\omega_d$. In all three cases we have fixed the following parameters; inner core $\alpha=0.1$, mean stratification $\bar{N}=1$, diffusivities $\nu=\kappa=10^{-6}$, and $Pr=1$.

\begin{figure}
\centering
	\subfigure[$\alpha=0.1$, $\beta=1.0$, $\bar{N}^2=1$, $\nu=\kappa=10^{-6}$, $\omega=0.1474$]{\label{fig:g-mode}
	\includegraphics[width=0.22\textwidth]{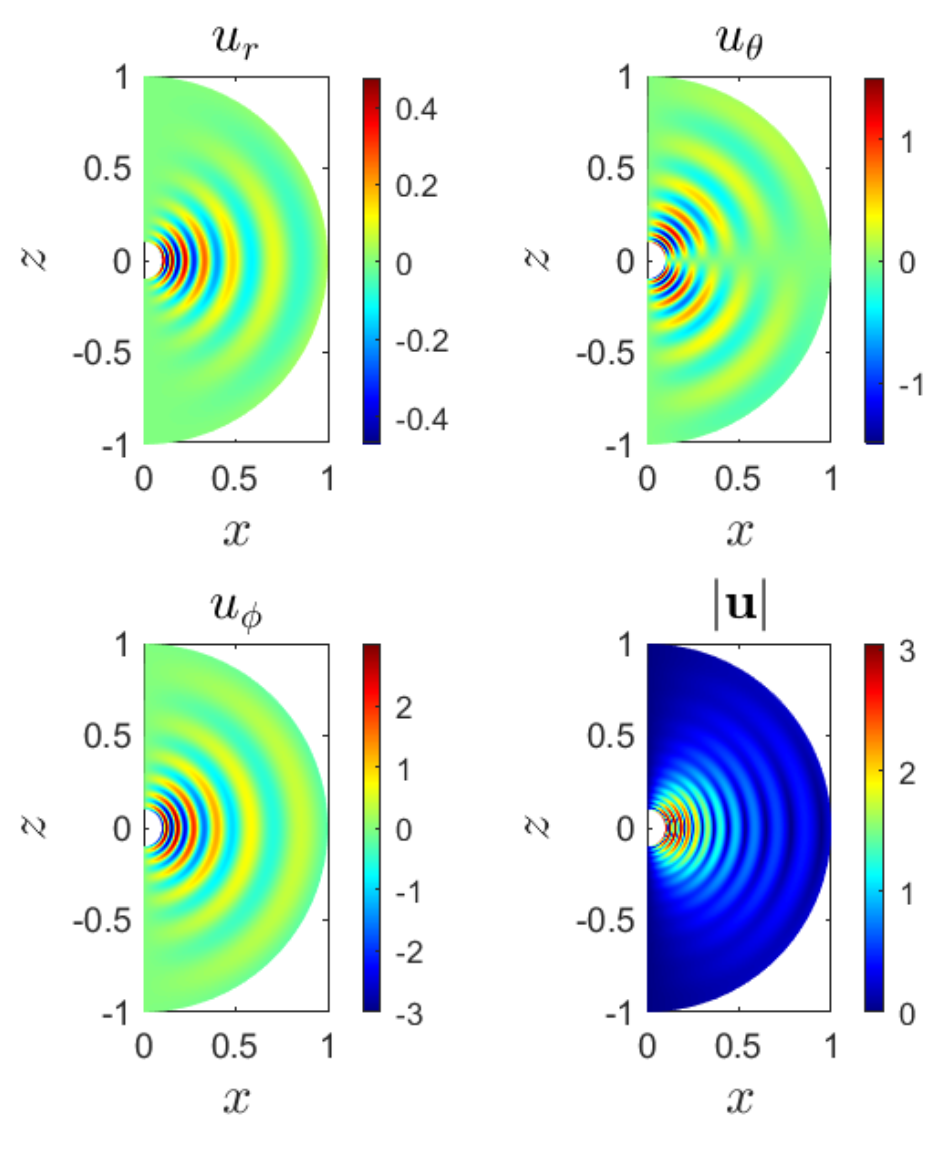}}
	\subfigure[$\alpha=0.1$, $\beta=1.0$, $\bar{N}^2=1$, $\nu=\kappa=10^{-6}$, $\omega=1.565$]{\label{fig:f-mode}
	\includegraphics[width=0.23\textwidth]{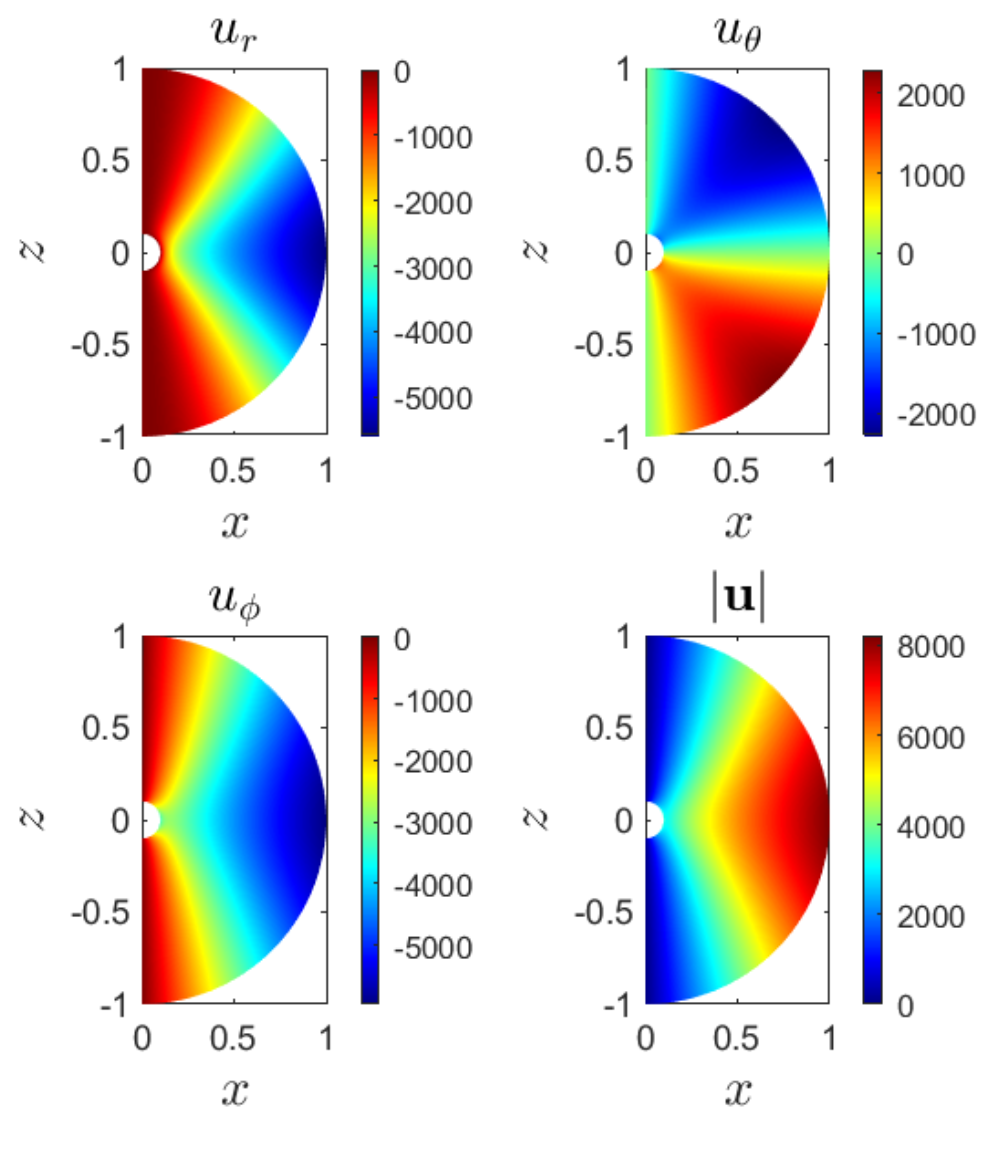}}
	\subfigure[$\alpha=0.1$, $\beta=0.5$, $\bar{N}^2=1$, steps$=3$, $\nu=\kappa=10^{-6}$, $\omega=0.4037$]{\label{fig:int-mode}
	\includegraphics[width=0.22\textwidth]{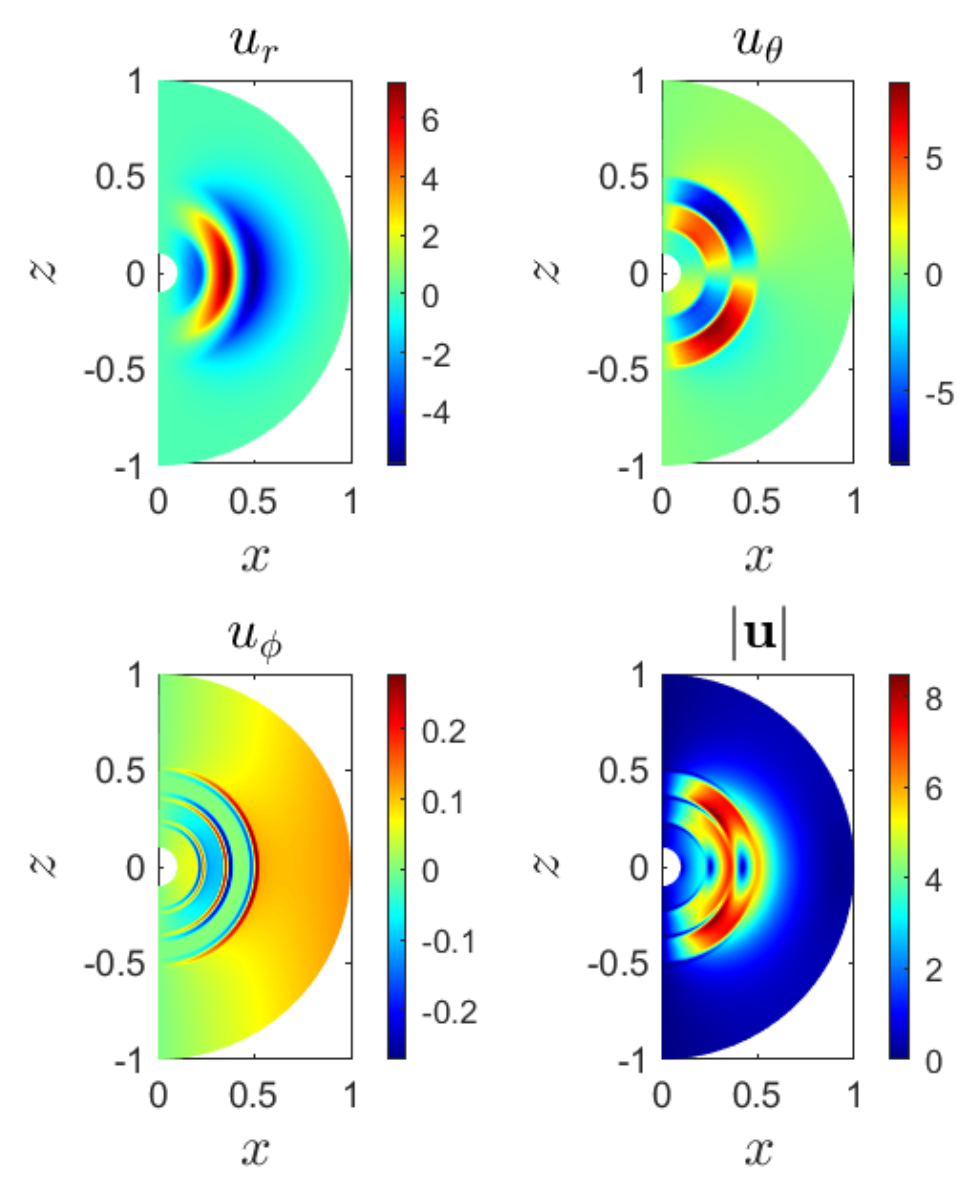}}
	\subfigure[$\alpha=0.1$, $\beta=0.5$, $\bar{N}^2=1$, steps$=3$, $\nu=\kappa=10^{-6}$, $\omega=0.09359$]{\label{fig:int_g-mode}
	\includegraphics[width=0.23\textwidth]{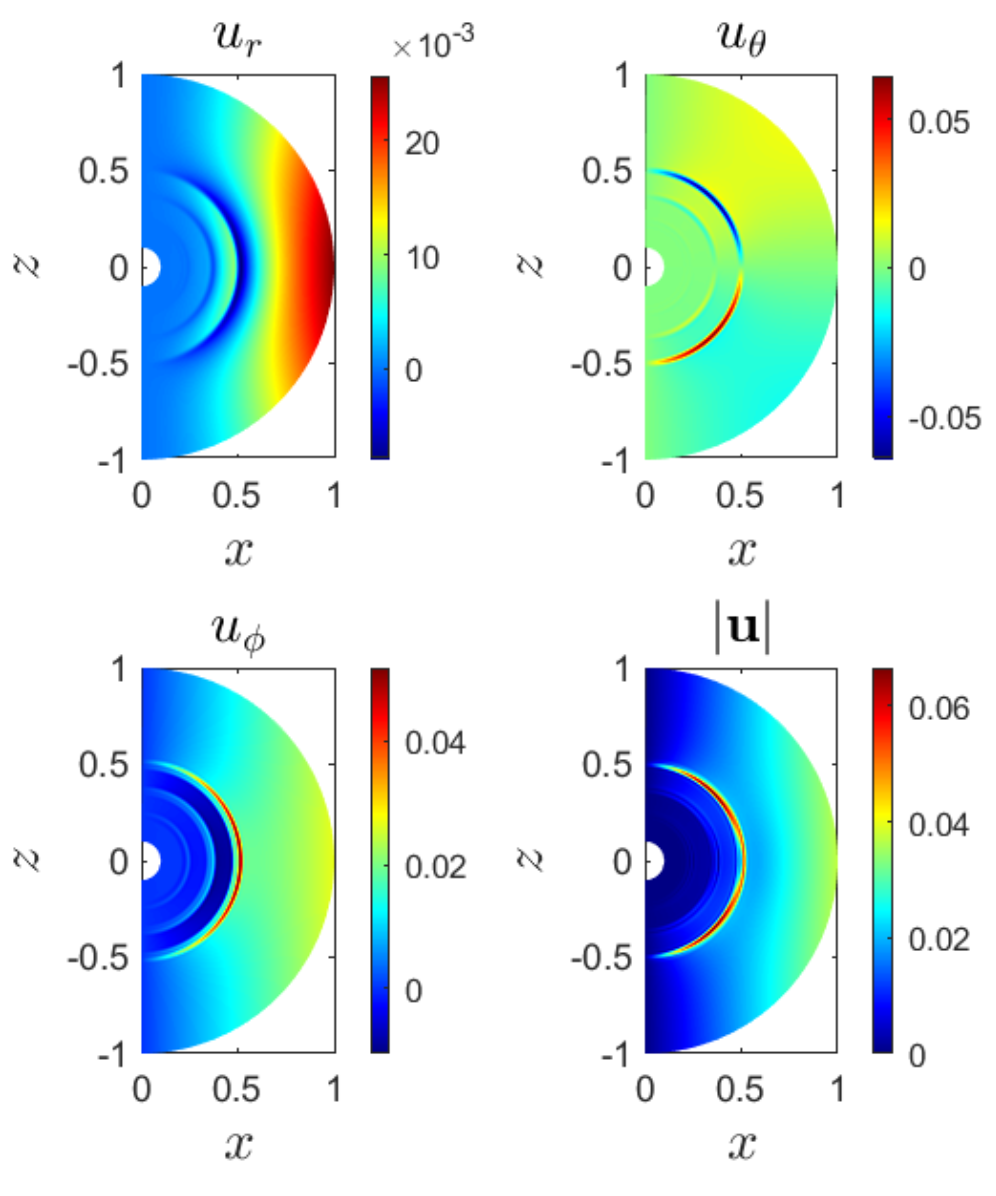}}
	\caption{Illustrative examples of the forced response with different stratification profiles and forcing frequencies, showing $u_r$, $u_{\theta}$, $u_{\phi}$ and $|\bf{u}|$. Top left: Example of an internal gravity wave (g-mode) propagation over the entire stratified layer. Top right: Example of a surface gravity (f-mode) response with the characteristic $Y^2_2$ shape. Bottom left: Example of interfacial modes where the staircase interfaces give a ``g-mode like" response. Bottom right: Example of g-mode response within the finite width of the staircase interface.} \label{fig:mode_over}
\end{figure}

Figure~\ref{fig:diss_over_a} is an example of a fully stratified interior, where a stably stratified layer with constant buoyancy frequency $\bar{N}=1$ extends from the solid inner core all the way to the planetary radius, described by equation \ref{eq:N2_layer} with $\beta=1$ and $\bar{N}=1$. Figure~\ref{fig:diss_over_b} similarly contains a uniformly stratified region, however in this case it extends to half the planetary radius ($\beta=0.5$). Finally, Figure~\ref{fig:diss_over_c} has a stratified layer that is equivalent to that shown in Figure~\ref{fig:diss_over_b} but with a staircase structure as described by equation \ref{eq:N2_steps} with $\beta=0.5$. In this case we consider also steps $=3$ and $\delta r=0.06$. In these (and all following non-rotating) cases we only show positive frequencies as the results are symmetric about $\omega=0$, $D(\omega)=D(-\omega)$, leading to no further information when considering negative frequencies. 

In all three cases we observe many tall narrow peaks of enhanced dissipation which occur over an extended frequency range. We note that the locations and magnitudes of these peaks vary considerably between these three examples. This is due to the strong link between the stratification profile and the properties of the modes, something we explore further in this section. This is carried out by examining the peaks and trends in dissipation, as well as comparisons to both the eigenvalue solutions from \S~\ref{sec:eigs} and analytical calculations for the modes from Appendices \ref{app:g-mode} and \ref{app:f-mode}. 

Our problem with uniform stratification is clearly very analogous to the tidal excitation of gravity waves in radiation zones of solar-type stars by short-period extrasolar planets or in stellar binary systems. In particular, prior work has studied the tidal excitation of global g-modes \citep[e.g.][]{Goodman1998,Terquem1998} in these stars, which share many qualitative properties with those in our model. In particular, the idea that there is efficient dissipation at specific g-mode resonant frequencies according to linear theory is exactly what we show here, and the idea of resonance locking in evolutionary scenarios has also been discussed \citep[e.g.][]{Witte2002}. In addition, the occurrence of a possible low-frequency travelling wave regime has also been discussed in these stars by \cite[e.g.][]{Goodman1998,Barker2010}, where it is primarily motivated by nonlinear effects such as wave breaking. We will elaborate on some of these connections further below.

\subsubsection{Uniform stable stratification extending to the planetary radius}

We look first at Figure \ref{fig:diss_over_a}, in which we are considering uniform stable stratification with $N^2=1$ throughout,  which allows for internal gravity waves (g-modes) to be excited and subsequently dissipated. These appear as a regular, discrete set of peaks, visible at frequencies less than $\bar{N}$, in agreement with the expected range for internal gravity waves. The purple squares are the frequencies found from the corresponding eigenvalue problem, where the $x$-axis is the frequency of oscillation $\operatorname{Re}[\omega]$, and the $y$-axis is the magnitude of the damping rate of the mode $|\operatorname{Im}[\omega]|$; note all modes are stable, as expected for $\bar{N}^2>0$. We see very good agreement between the eigenvalue solution and the sharp resonant peak and a clear increase in damping rate as the forcing frequency decreases. This corresponds with the clear decrease in peak height and increase in width, in agreement with our expectations based on the forced damped simple harmonic oscillator discussed in \S~\ref{sec:SHO}. The green crosses which represent the analytical prediction of the internal gravity mode frequencies also agree well with the dissipation peaks. The slight discrepancies are expected due to neglected factors in the analytical calculation (viscosity, thermal diffusivity, and the departure from the free-surface condition by the use of solid-wall boundary conditions). The discrepancies are more significant at higher frequencies, although these solutions have a larger characteristic wavelength and are therefore less affected by viscosity; they are more affected by the free-surface condition, which has a larger contribution to shifting the mode frequency. In this case we do not have a straightforward value for the damping rate as dissipation was neglected to allow for analytical solutions, therefore the $y-$axis value is arbitrary. 

To understand the mode properties further we look at the spatial structure of the solutions at a given frequency. Figure \ref{fig:g-mode} shows the forced response for the uniform case shown in Figure \ref{fig:diss_over_a} at a frequency of $\omega=0.1474$, well within the range characteristic of internal gravity waves ($\omega < \bar{N}$). We observe a series of oscillations in $r$, characteristic of an internal gravity wave, where the number of nodes (and corresponding wavelength) is frequency-dependent, with wavelength decreasing with decreasing frequency. Figure \ref{fig:f-mode} shows the solution at $\omega=1.565$, also for the model shown in Figure \ref{fig:diss_over_a}. This is the surface gravity (f-mode) response, introduced because we are using the free-surface boundary condition at the planetary surface. It corresponds to the large peak around $\omega=1.5$, approximately $\omega=\omega_d \sqrt{\ell}$ with $l=2$. This is the expected location of the surface gravity mode for a homogeneous body, neglecting self-gravity, with $l=2$ \citep{Barker2016}. The mode has been shifted to a higher frequency due to effects of both the finite core size and stable stratification. Analytical approximations incorporating both of these shifts are provided in Appendix \ref{app:f-mode} and are plotted on Figure \ref{fig:diss_over_a}.
The shift due to a stably stratified interior is shown by the blue circle and can be seen to agree well with the resulting resonance; the shift due to a finite core size in this case is far less significant than the shift due to stratification and therefore can be neglected. The shape of this response for $u_r$ is characteristic of a $Y^2_2$ spherical harmonic shape, which can be identified by the single zero crossing in $\theta$, as the number of zero crossings is equal to $l-|m|$. 

\subsubsection{Uniform stable stratification extending to half the planetary radius}

Figure \ref{fig:diss_over_b}, the case where uniform stratification extends to half the planetary radius, has many similar properties to Figure \ref{fig:diss_over_a}. We again see the regular, discrete set of peaks at frequencies less than the buoyancy frequency, $\omega < \bar{N}$, corresponding to internal gravity modes, and a resonance with the surface gravity mode around $\sqrt{2}$, all of which agree well with the numerical eigenvalues. However, we can see a number of differences, most notably in our agreement with analytical calculations. The agreement with the analytical calculation of the internal gravity modes (Appendix \ref{app:g-mode}) is worse. As we are now considering a stratified layer beneath a convective medium, the impact of solid-wall boundary conditions is more significant. The frequency of the surface gravity mode agrees more closely with the predicted shift due to a finite core than that of a stratified planet, which is to be expected because our analytical calculation neglects the convective envelope. Finally we note that the overall magnitude of dissipation is lower; this will be discussed further in \S~\ref{sec:param_size}. Further analysis of the spatial structure shows similar behaviour to the previous case so will be omitted.

\subsubsection{Staircase structure extending to half the planetary radius}

Figure \ref{fig:diss_over_c} shows the dissipation in a system with a semi-convective layer where a staircase-like density profile sits above the core, extending to half the planetary radius, above which we have a convective region. This is relevant as a staircase is a potential result of double-diffusive convection in giant planet interiors. As in the first two examples, we see good agreement with the eigenvalue solutions (purple squares) in both oscillatory frequency and damping rate.

The staircase we are considering has three steps and we observe three clear corresponding peaks. These peaks align with three internal wave frequencies of the system which correspond to the modes found and discussed in \cite{Pontin2020}. Figure \ref{fig:int-mode} shows the forced solution at $\omega=0.4037$, where the adjacent interfaces can be seen to be oscillating out of phase with one another, exhibiting ``g-mode like" behaviour. This mode is analogous to that of an internal gravity mode with three interior nodes. The green crosses show the predicted values for an equivalent uniformly stratified layer, and we can see that although the interfacial mode peaks appear in the same frequency range as the internal gravity modes $\omega < \bar{N}$, the exact frequency of the peaks has been shifted due to the properties of the staircase. 

Additionally in Figure \ref{fig:diss_over_c}, we can see a small collection of resonances at low frequencies, approximately near $\omega=0.1$, an example of which is shown in Figure \ref{fig:int_g-mode}. These are internal gravity wave modes sustained by the finite width of the interfaces. These small regions of stable stratification allow internal modes to exist, which in turn allow wave resonances at low frequencies. The blue crosses predict the internal gravity mode resonances expected to form within the finite width of the interface (Appendix \ref{app:g-mode}), and can be seen to correspond to the additional peaks observed. We expect slightly different frequencies for each of the three steps, however here we plot just one set for clarity.

The surface gravity (f-mode) response is again dominant in Figure \ref{fig:diss_over_c}, however it is no longer significantly affected by the stably stratified layer. The shift due to a finite core at $\alpha=0.1$ (red circle) predicts the location of the peak well, but the adjustment predicted by stratification does not, due to its neglect of the convective region from $\beta r_0$ to $r_0$. 

\subsection{Low frequency, travelling wave regime}\label{sec:trav_wave}

To aid understanding of our numerical results we first consider an analytical limit of our model. We assume that the damping mechanism is efficient such that a wave excited at the outer boundary that propagates inwards is fully damped before reflecting from the core and forming a standing wave. Therefore, we can assume that all the energy from the wave is dissipated and equate the inward travelling energy flux with the total tidal dissipation rate. In this model our damping mechanism is kinematic viscosity and thermal diffusivity, and therefore this regime applies to low frequency waves which have sufficiently short wavelengths. 

For this approximation to hold, we consider frequencies for which the damping timescale is shorter than the group travel time. Details of the calculation are found in Appendix~\ref{app:trav_wave}, where the critical frequency is found to be approximately
\begin{equation}\label{eq:omega_crit}
\omega_{crit} = \big((1-\alpha) r_0 (\nu+\kappa) (N k_{\perp})^3 \big) ^{\frac{1}{4}}. 
\end{equation}
Although this approximation only considers low frequencies, this regime is potentially relevant for planetary applications if the forcing frequency is sufficiently low compared with the dynamical frequency of the planet, provided waves are fully damped. This calculation is independent of the damping mechanism so it also holds when considering other mechanisms such as wave breaking or critical layer absorption, provided nearly complete absorption occurs \citep{Barker2010,Su2020}.

\begin{figure}
	\centering
	\subfigure[Radial dependence of forced solution to illustrate the propagation depth and critical frequency, for $\alpha=0.1$, $\beta=1.0$, $\bar{N}=1$, $\omega=0.05$.]{\includegraphics[width=0.46\textwidth]{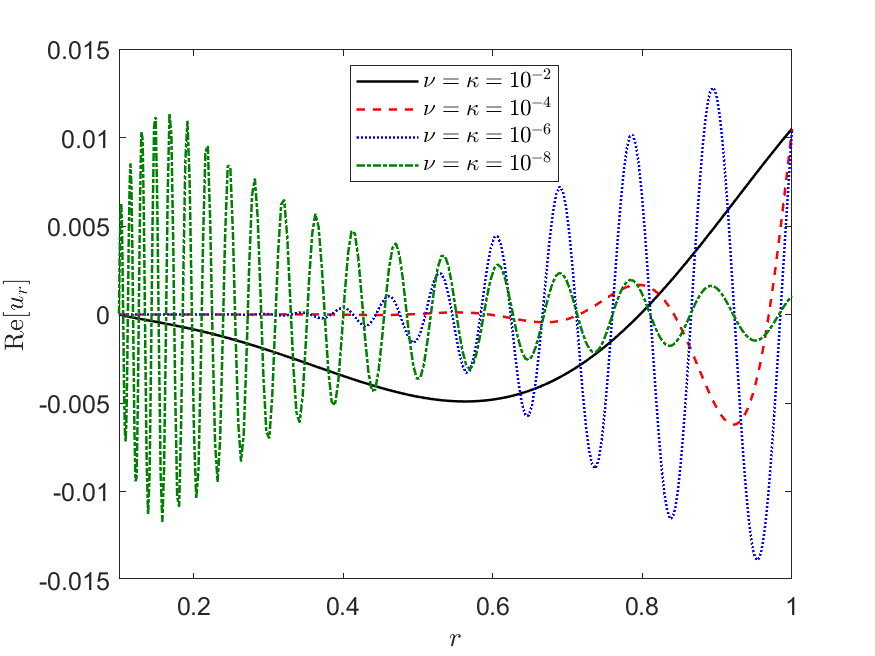}
\label{fig:trav_wave_sol}}
\subfigure[Total dissipation for analytical travelling wave calculation (crosses and straight lines) and numerical result (solid line). Vertical lines show the critical frequency given by equation~\ref{eq:omega_crit}.]{\includegraphics[width=0.46\textwidth]{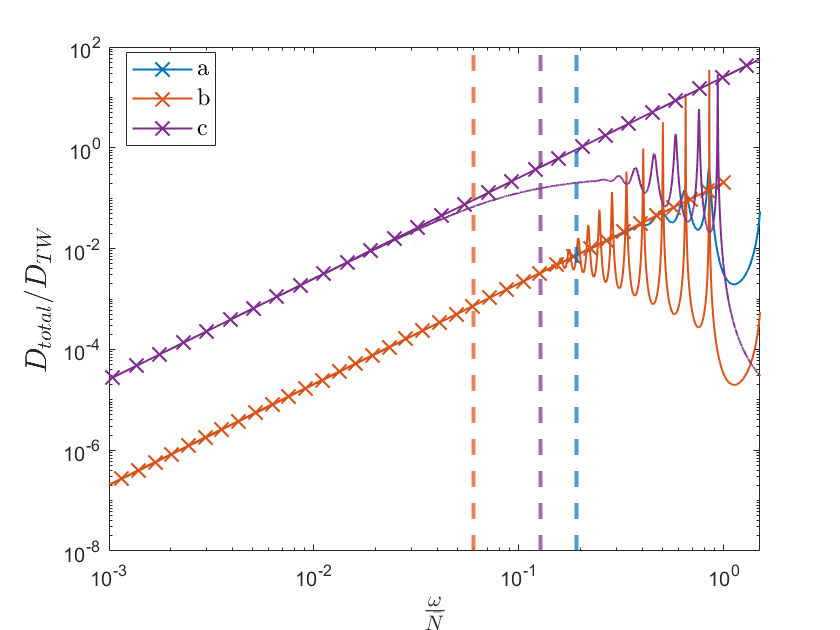} \label{fig:trav_wave}}
\subfigure[Comparison of the analytical travelling wave calculation and the numerical frequency-averaged dissipation using $\omega_{max}=\omega_{crit}$.]{\includegraphics[width=0.46\textwidth]{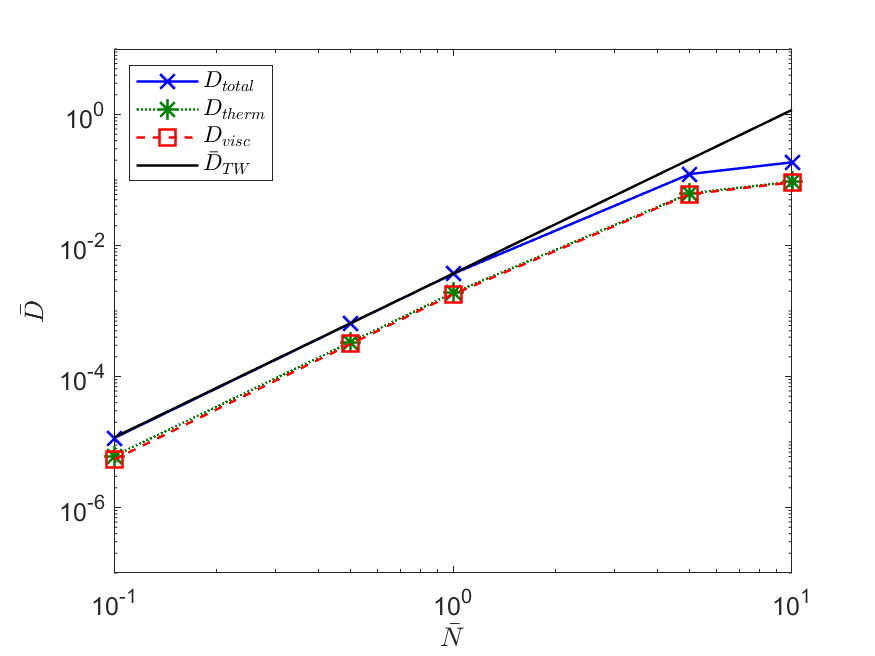}\label{fig:trav_wave_int}}
	\caption{
	Comparison of the travelling wave calculations and numerical results, for all cases $\alpha=0.1$, $\beta=1$ and in panel \protect\subref{fig:trav_wave_int} $\nu=\kappa=10^{-4}$. }\label{fig:trav_wave_over}
\end{figure}

\begin{figure}
	\centering
	\subfigure[$\nu=\kappa=10^{-4}$]{\includegraphics[width=0.46\textwidth]{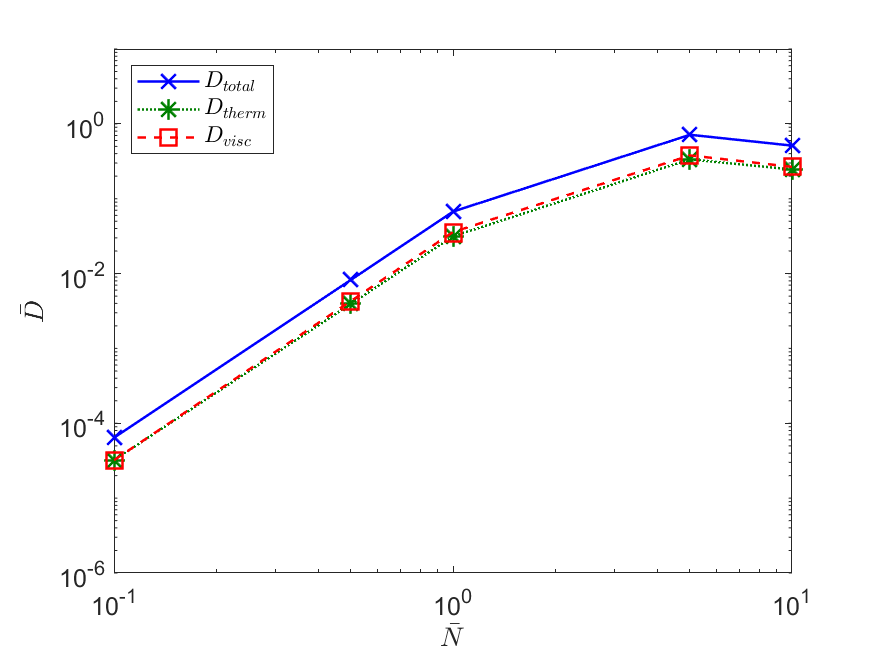}\label{fig:diss_N_nu4_int}}
	\subfigure[$\nu=\kappa=10^{-6}$]{\includegraphics[width=0.46\textwidth]{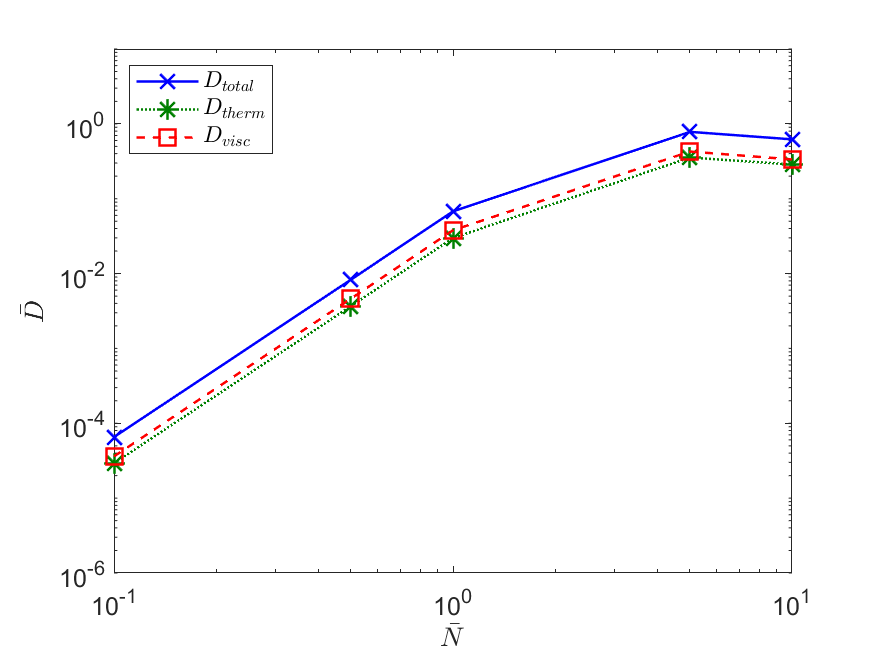}\label{fig:diss_N_nu6_int}}
	\caption{Frequency-averaged dissipation with $\frac{1}{\omega}$ weighting, as a function of stratification strength $\bar{N}$ for two different viscosities and thermal diffusivities. In both cases $\omega_{max}=\bar{N}$ and $\alpha=0.1$, $\beta=1.0$. } \label{fig:diss_N_int}
\end{figure}

Figure \ref{fig:trav_wave_sol} shows the radial dependence of the solution at a forcing frequency of $\omega=0.05$ for various different viscosity and thermal diffusivity values, with other parameters kept fixed at $\alpha=0.1$, $\beta=1$, $\bar{N}=1$ and $Pr=1$. We can see that as viscosity increases, as well as altering the wavelength of the solution (see \S~\ref{sec:param_visc}), the depth within the planet to which the wave propagates before being absorbed is decreased. 
We see that for both the $\nu=10^{-2}$ (black solid) and $\nu=10^{-4}$ (red dashed) cases, which are well below the critical frequency (0.0717 and 0.0227 respectively) at which the waves are absorbed before reaching the inner boundary. For the $\nu=10^{-4}$ (blue dotted) case where we are approaching the critical frequency (0.0717) the wave propagates most of the way through the interior. Finally in the $\nu=10^{-8}$ (green dot-dashed) case we are above the critical frequency (0.0227) and the wave behaviour extends throughout the entire domain. 

As complete wave absorption has occurred, we only need to consider the inwardly travelling component of the wave solution, setting the outward coefficient to be identically zero. This allows us to calculate the total energy flux in the wave and equate it to the total dissipation. In Appendix~\ref{app:trav_wave}, we show that if $\beta=1$, the total dissipation is then described by
\begin{equation}\label{eq:D_TV}
D_{TW}= \frac{|\psi_0|^2 r_0^{5} \bar{N} \omega^2}{2 \sqrt{l(l+1)}}.
\end{equation}

Figure \ref{fig:trav_wave} shows the analytical dissipation as described by equation \ref{eq:D_TV}, compared to numerical results for three cases. All three cases have $\alpha=0.1$, $\beta=1$, and the blue, red and yellow lines are for $\bar{N}=1$ and $\nu=\kappa=10^{-4}$, $\bar{N}=1$ and $\nu=\kappa=10^{-6}$, and $\bar{N}=5$ and $\nu=\kappa=10^{-4}$, respectively. We can see when comparing at low frequencies there is particularly good quantitative agreement between the travelling wave calculation (crosses) and numerical results (solid lines), where both follow a clear $\omega^2$ trend with the same slope. The dashed lines indicate the critical frequencies below which this approximation would be valid, and we can see that this estimation approximately holds. 

At sufficiently low frequencies we observe no dependence on viscosity or thermal diffusivity as both the blue and red lines agree. As we assume the wave is fully damped, the timescale over which dissipation occurs is not then significant. By comparing the blue and yellow lines, we can see that there is a linear dependence on $\bar{N}$, showing the strength of the stable stratification does play an important role in the magnitude of dissipation. 

A similar regime for inward propagating travelling waves has been studied in solar-type stars \citep[e.g.][]{Goodman1998,Barker2010} where nonlinear effects have been proposed to be responsible. One important difference is that in solar-type stars frequencies of tidal forcing are often low enough that short-wavelength travelling waves are thought to be launched from a narrow region in the vicinity of the radiative/convective interface. The local radial profile of $N^2$ in this region affects the propagating energy flux \citep[e.g.][]{Barker2011}. Here we have considered a constant $N^2$ in the stable layer for simplicity, but we note that the energy flux in lower frequency gravity waves would be affected by the spatial variation in $N^2$ near the interface with any overlying convection zone.

In Appendix \ref{app:trav_wave} we also find the frequency-averaged (with a $\frac{1}{\omega}$ weighting) travelling wave dissipation defined by equation \ref{eq:freq_avg}. When we take limits of $\omega_{min}=0$ and \hbox{$\omega_{max}=\bar{N}$,} it is found that
\begin{equation}\label{eq:trav_N}
\bar{D}_{TW}=\frac{|\psi_0|^2 r_0^{5} \bar{N}^3}{4 \sqrt{l(l+1)}},
\end{equation}
and when we take $\omega_{max}=\omega_{crit}$ and $\omega_{min}=0$, we find
\begin{equation}
\bar{D}_{TW}=\frac{|\psi_0|^2 r_0^{5} \bar{N}^{\frac{5}{2}} (\beta-\alpha)^{\frac{1}{2}} (\nu +\kappa)^{\frac{1}{2}}  (l(l+1))^{\frac{1}{4}}}{4}.
\end{equation}

We compare the travelling wave frequency-averaged dissipation to the numerically equivalent calculation, using $\omega_{crit}$ as our upper integration bound in both cases. Figure \ref{fig:trav_wave_int} shows that the numerical calculation agrees well with the travelling wave approximation, where we have $\alpha=0.1$, $\beta=1$, $\nu=\kappa=10^{-4}$ and we vary $\bar{N}$. We can see that for small values the agreement is almost exact, and only begins to deviate for $\bar{N} > 1$. This agrees with the departure between analytical and numerical results seen in Figure \ref{fig:trav_wave}, and is a result of the relevant damping per wavelength decreasing as we increase $\bar{N}$, to be discussed further in \S~\ref{sec:param_N}. 

\subsection{Parameter dependencies for uniform stable stratification}\label{sec:params}

We now consider how the key parameters in our model alter the expected viscous, thermal and total dissipation, by varying each parameter in turn. 

\begin{figure*}
	\subfigure[$\nu=\kappa=10^{-4}$]{\includegraphics[width=0.49\textwidth]{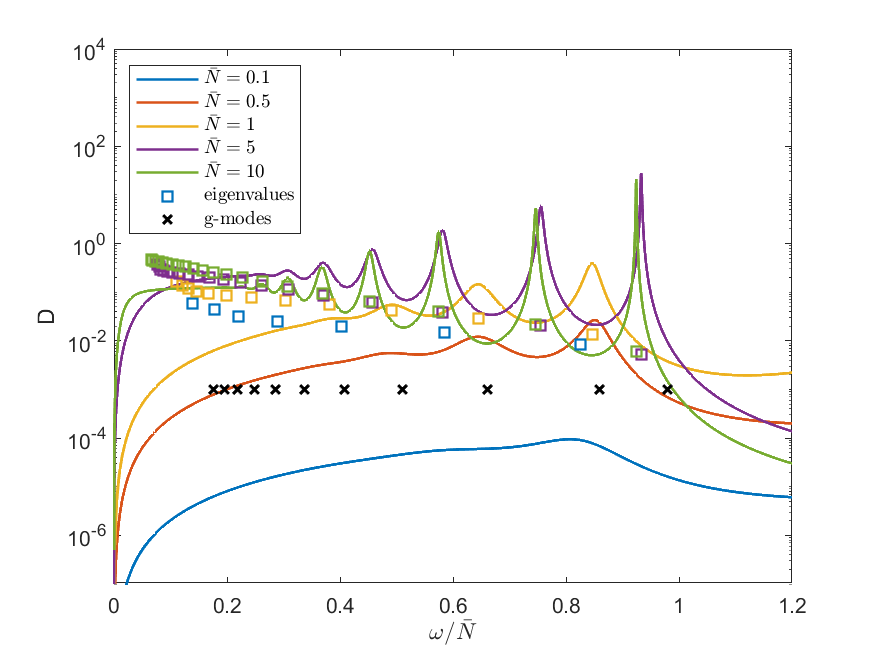}\label{fig:diss_N_4}}
	\subfigure[$\nu=\kappa=10^{-4}$]{\includegraphics[width=0.49\textwidth]{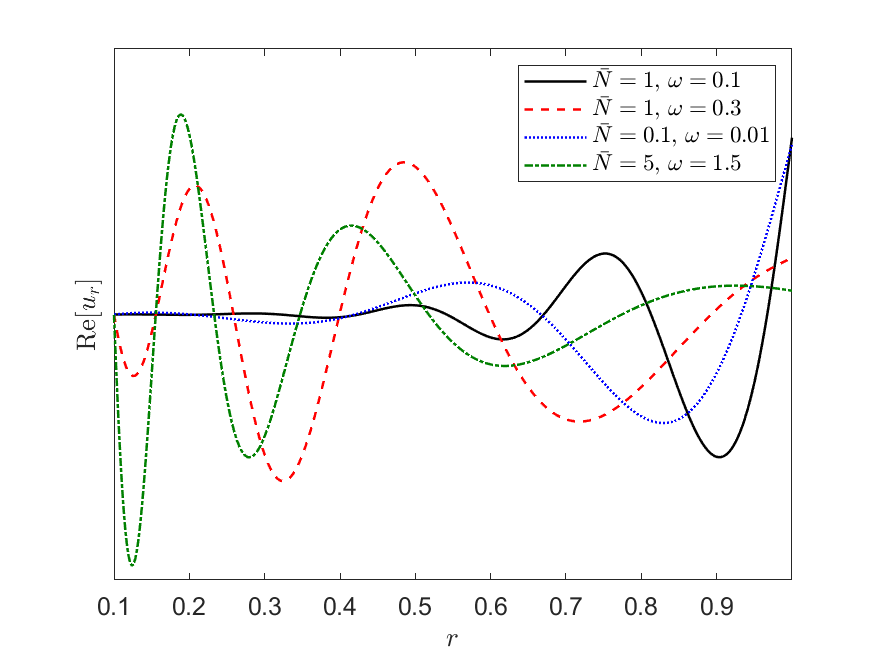}\label{fig:omega_N_4}}
	\subfigure[$\nu=\kappa=10^{-6}$]{\includegraphics[width=0.49\textwidth]{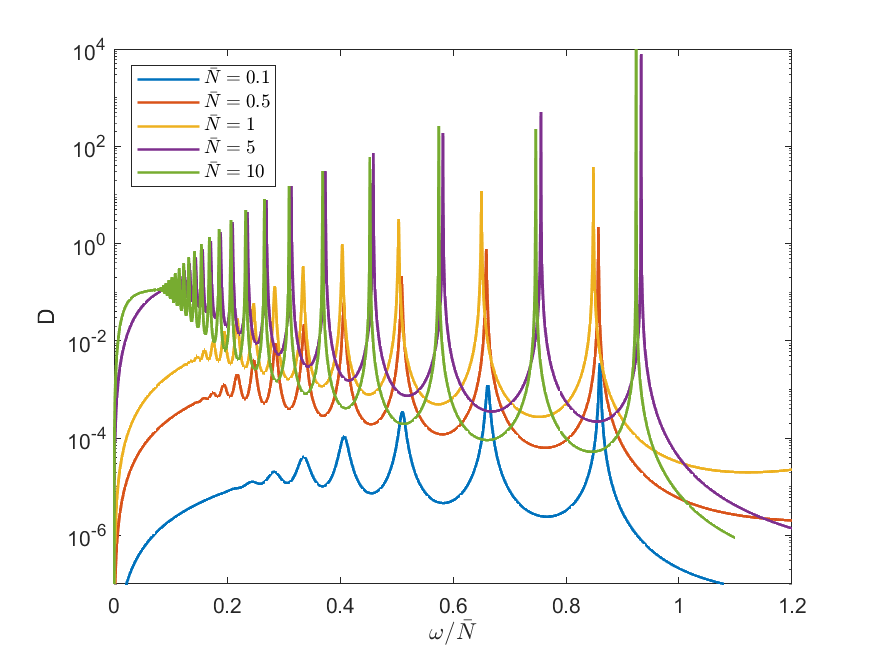}\label{fig:diss_N_6}}
	\subfigure[$\nu=\kappa=10^{-6}$]{\includegraphics[width=0.49\textwidth]{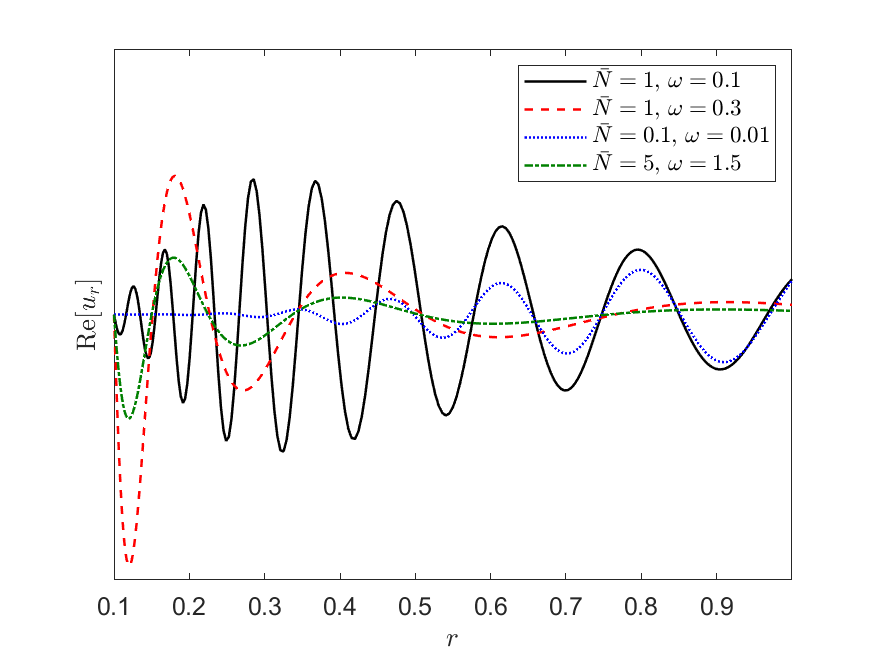}\label{fig:omega_N_6}}
	\caption{Panels \protect\subref{fig:diss_N_4} and \protect\subref{fig:diss_N_6} show dissipation rate as a function of frequency for different magnitudes of uniform stable stratification. Panels \protect\subref{fig:omega_N_4} and \protect\subref{fig:omega_N_6} show the radial dependence of $u_r$ for the given parameters. Note the $y-$axis has been scaled for illustrative purposes, $(\bar{N}=1, \omega=0.1) \times 50$, $(\bar{N}=1, \omega=0.3) \times 5$, $(\bar{N}=0.1, \omega=0.01) \times 500$.} \label{fig:N_6_over}
\end{figure*}

\subsubsection{Varying the strength of the stable stratification}\label{sec:param_N}

The key parameter for the stable stratification is the buoyancy frequency, $\bar{N}$, which is a measure of the strength of the stratification as it dictates the density/entropy gradient. Planetary evolution models and observations give a range of results for the properties of the stable stratification, therefore here we explore what implications different values could have.  

Figure \ref{fig:diss_N_int} shows the frequency-averaged dissipation as a function of $\bar{N}$ showing the viscous $\bar{D}_{visc}$, thermal $\bar{D}_{ther}$, and total $\bar{D}_{total}$ frequency-averages, where integration limits have been taken as $\omega_{min}=0$ and $\omega_{max}=\bar{N}$. In both cases we are considering a uniform stable stratification extending to the planetary radius, therefore $\alpha=0.1$ and $\beta=1$, however we consider two different values for the viscosity and thermal diffusivity, such that Figure~\ref{fig:diss_N_nu4_int} is for the case $\nu=\kappa=10^{-4}$ and Figure~\ref{fig:diss_N_nu6_int} for $\nu=\kappa=10^{-6}$.  At this stage we simply note the similarity between the two plots and discuss them concurrently, expanding on the similarities further in \S~\ref{sec:param_visc}. There is a clear $\bar{N}^3$ dependence when $\bar{N}<1$, which agrees with the analytical travelling wave calculation given by equation \ref{eq:trav_N} discussed previously. This holds despite the change in the limit used for $\omega_{max}$, as in this case we have integrated beyond the critical frequency so would not necessarily expect complete agreement. We note that the same trend is exhibited in both the thermal and viscous dissipation as the resonant waves dissipate through both mechanisms. The viscous and thermal dissipation rates are similar to within approximately a factor of 2 here presumably because $\nu=\kappa$ (we vary these parameters separately in \S~\ref{sec:param_visc}).

\begin{figure}
	\centering
\subfigure{
	\begin{tikzpicture}
		\node at (0,0) {\includegraphics[width=0.46\textwidth]{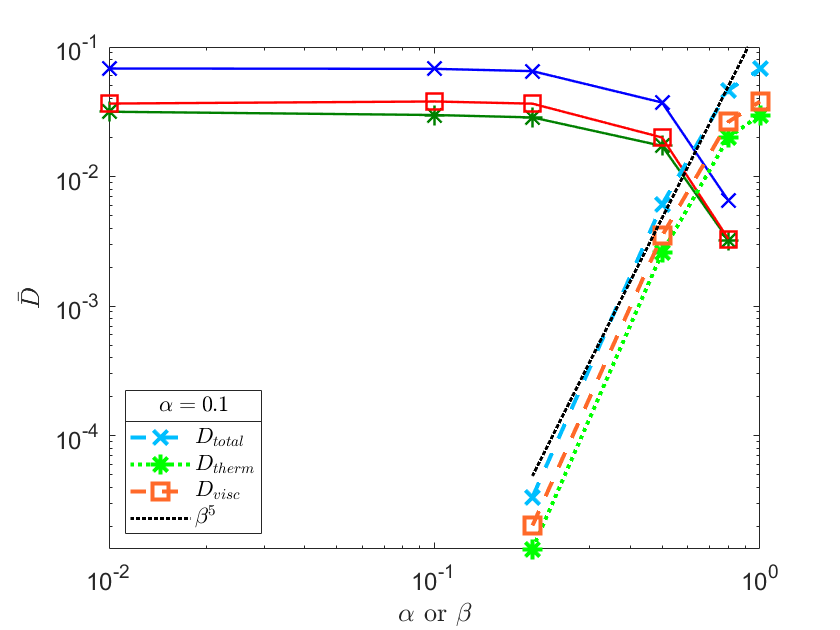}};
		\node at (-2.15,0.35) {\includegraphics[width=0.08\textwidth]{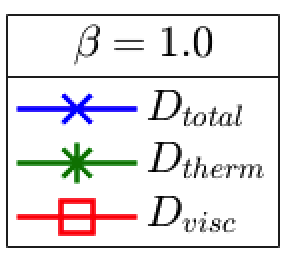}};
	\end{tikzpicture}
\label{fig:diss_ab_loglog}}
	\subfigure{
	\begin{tikzpicture}
		\node at (0,0) {\includegraphics[width=0.46\textwidth]{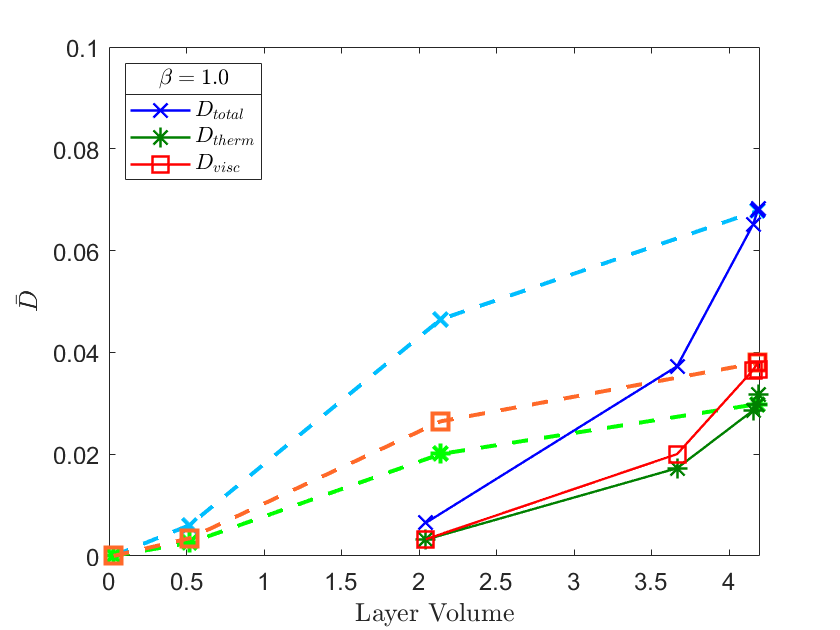}};
		\node at (2.5,1.9) {\includegraphics[width=0.09\textwidth]{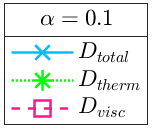}};
	\end{tikzpicture}
	\label{fig:diss_vol_int}}
	\caption{Frequency-averaged dissipation as a function of stratified layer size and volume, varying $\alpha$ and $\beta$. In all cases $\omega_{max}=\bar{N}$ and other parameters are fixed, $\bar{N}=1$, $\nu=\kappa=10^{-6}$. } \label{fig:diss_size_int}
\end{figure}

\begin{figure}
	\centering
	\subfigure{\includegraphics[width=0.46\textwidth]{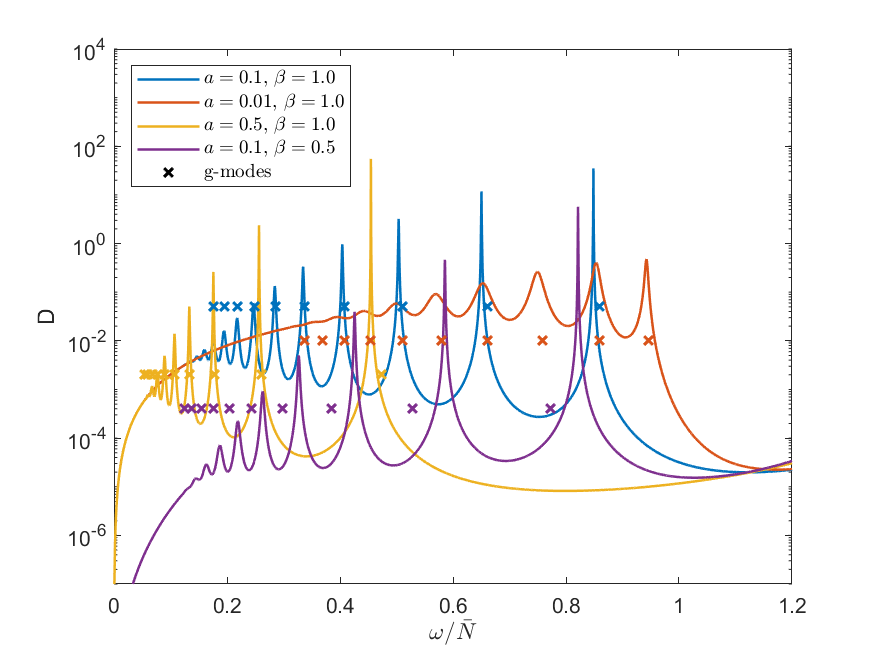}\label{fig:diss_ab}}
	\subfigure{\includegraphics[width=0.46\textwidth]{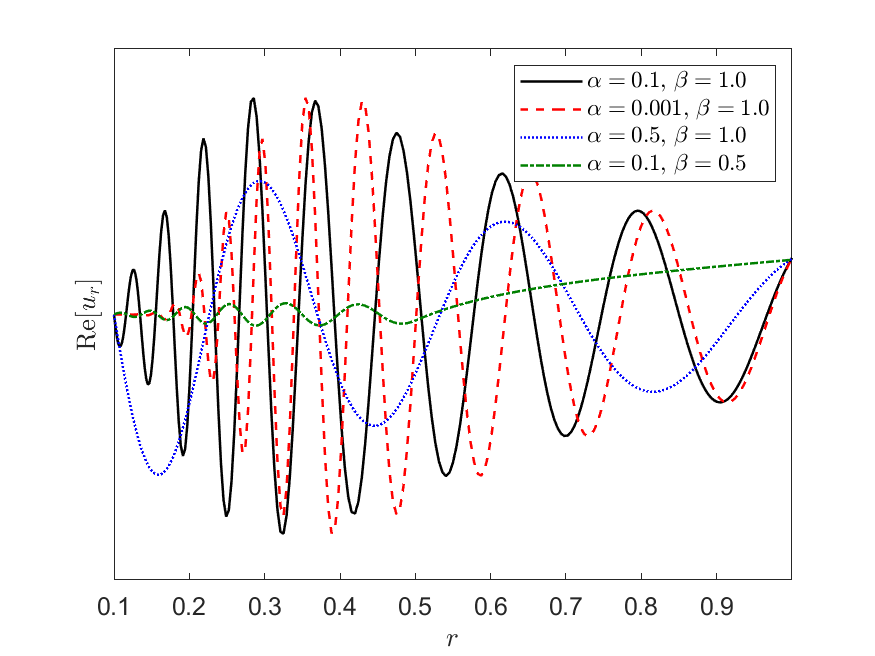}\label{fig:omega_ab}}
		\caption{Panel \protect\subref{fig:diss_ab} shows total dissipation rate as a function of frequency for various $\alpha$ and $\beta$ values. Panel \protect\subref{fig:omega_ab} shows the radial dependence of $u_r$ for given parameters.} \label{fig:ab_over}
\end{figure}

Figures \ref{fig:diss_N_4} and \ref{fig:diss_N_6} show the frequency dependence of dissipation for the cases, which when integrated, give the data for the frequency-averaged values shown in Figures \ref{fig:diss_N_nu4_int} and \ref{fig:diss_N_nu6_int} respectively. The $x$-axis has been scaled by $\bar{N}$ in these figures to allow for easy comparison. We can clearly see the characteristic g-mode peaks that appear at a similar location in $\frac{\omega}{\bar{N}}$ in each case. The first mode, located just below $\bar{N}$, contains one node in radius, therefore the wavelength is equal to the layer depth. Following on from this, additional lower frequency modes correspond to waves with additional nodes, which have lower frequencies and shorter wavelengths. These exist as an infinite set of modes, see equation \ref{eq:g_ana}, however the first 10 have been plotted in this case.  The frequency of the internal gravity modes obtained analytically is linear in $\bar{N}$, therefore when plotting on the $\frac{\omega}{\bar{N}}$ axis the predictions lie on top of one another.  The modes of the full system clearly have a dependence on $\bar{N}$, and roughly speaking the $\bar{N}<1$ trends deviate more than $\bar{N}>1$ cases. There are several factors in balance here; the effect of neglecting viscosity and thermal diffusivity is more prevalent at low frequencies and low $\bar{N}$, and the effect of the free surface is more prevalent at frequencies comparable with $\omega_d$ and large $\bar{N}$. It is the balance between these opposing effects that gives the non-obvious dependence in accuracy.

As $\bar{N}$ increases, more resonances are visible as clear peaks and the modes appear to be narrower and sharper. Though this is initially counter-intuitive, as the eigenvalue solutions, shown by the squares with the corresponding colour, show that the damping rate increases with $\bar{N}$, it can be explained by considering the relevant timescales.  If we consider the ratio of the group travel time to the damping time, we find the following,
\begin{equation}
\frac{t_{g}}{t_{d}}=\frac{2(\beta-\alpha) r_0 k_r^2}{k_{\perp} \bar{N}} \frac{(\nu + \kappa) k_r^2}{2} 
 \approx \frac{ (\nu + \kappa)  (\beta-\alpha) r_0}{\bar{N}}\frac{\bar{N}^4k_\perp^4}{\omega^4},
\end{equation} 
since $\omega^2\approx \bar{N}^2 k_\perp^2/k_r^2$ for low frequencies. We can see that the ratio $t_d/t_g$ is linearly dependent on $\bar{N}$ for fixed ratios of $\omega/\bar{N}$ and therefore despite the overall damping rate ($\propto t_d^{-1}$) increasing the “damping per wave crossing time for fixed $\omega/\bar{N}$" decreases which leads to sharper and narrower peaks (though there is a larger $t_g/t_d$ for larger $\bar{N}$ at fixed $\omega$).

Figures \ref{fig:omega_N_4} and \ref{fig:omega_N_6} show the forced solution for $u_r$ as a function of $r$ for different frequencies and stratification strengths, but note the $y$-axis has been scaled so all examples can be viewed on one plot.  All the cases shown are examples of internal gravity (g-mode) waves as shown in Figure \ref{fig:g-mode}. Both the wavelength and amplitude of the solutions are strongly dependent on the forcing frequency and stratification. The amplitude (not shown on these plots), depends on how close the forcing frequency is to a resonance, where we note that dissipation is proportional to amplitude squared. 

\begin{figure}
	\centering
	\subfigure[$Pr=1$]{\includegraphics[width=0.45\textwidth]{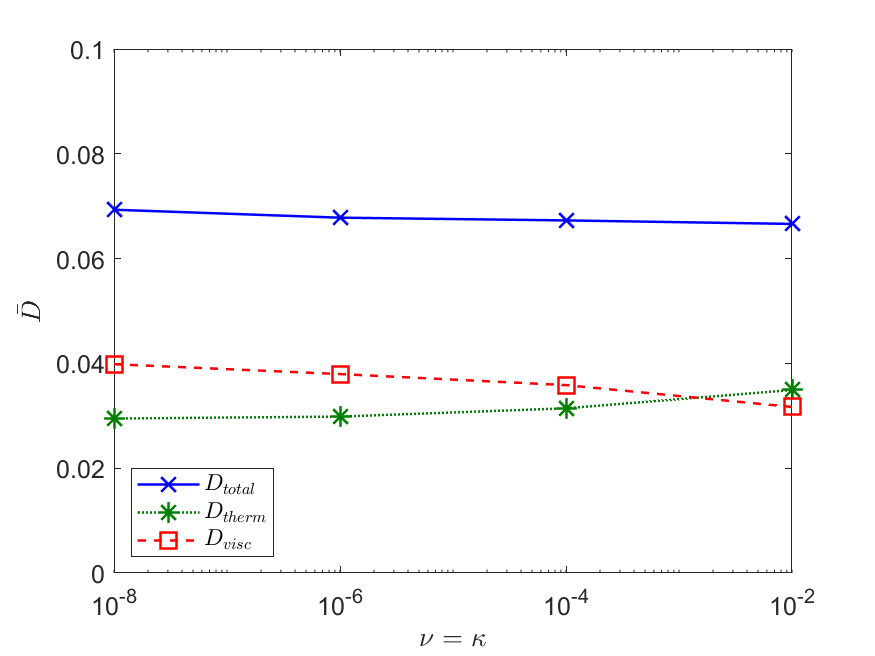}\label{fig:diss_nu_int}}
	\subfigure[$\kappa=10^{-2}$]{\includegraphics[width=0.45\textwidth]{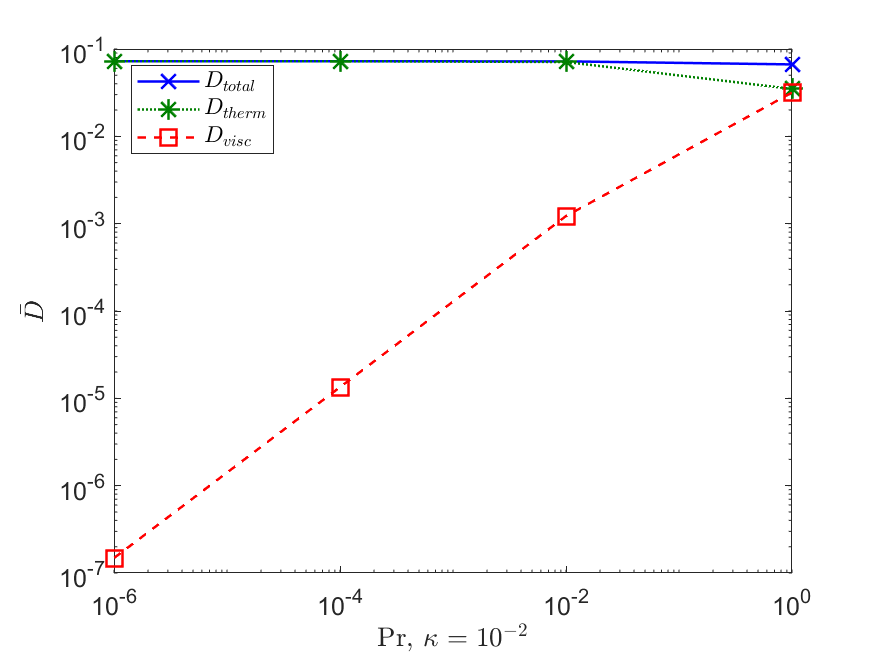}\label{fig:diss_Pr_nu_int}}
	\subfigure[$\nu=10^{-8}$]{\includegraphics[width=0.45\textwidth]{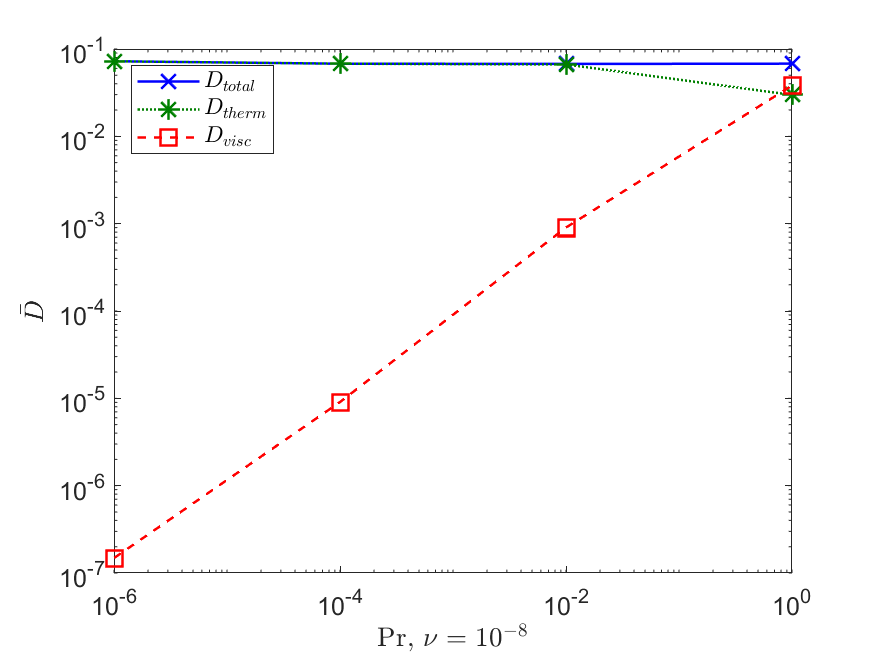}\label{fig:diss_Pr_kappa_int}}
	\caption{Frequency-averaged dissipation as a function of viscosity, thermal diffusivity and Prandtl number, for $\alpha=0.1$, $\beta=1.0$, $\bar{N}=1$. In all cases $\omega_{max}=\bar{N}$. } \label{fig:diss_visc_int}
\end{figure}

By comparing the solid black ($\bar{N}=1$, $\omega=0.1$) and red dashed ($\bar{N}=1$, $\omega=0.3$) lines we can see that decreasing the forcing frequency (i.e. ratio of $\frac{\omega}{\bar{N}}$), decreases the wavelength. Each peak in dissipation corresponds to a resonance where a multiple of half-integer wavelengths fits within the stratified layer. This relationship can be more clearly seen in the $\nu=\kappa=10^{-6}$ case as the frequencies shown are both well above the critical frequency ($\omega_{crit}=0.072$).  By comparing the black solid line ($\bar{N}=1$, $\omega=0.1$) with the blue dotted line ($\bar{N}=0.1$, $\omega=0.01$) and the red dashed line ($\bar{N}=1$, $\omega=0.3$) with the  green dot-dashed line ($\bar{N}=5$, $\omega=1.5$) we can see that the wavelengths are roughly comparable for the same $\frac{\omega}{\bar{N}}$. Again, proximity to the critical frequency affects the robustness of this result. 

\subsubsection{Varying the size of the stably stratified layer} \label{sec:param_size}

In realistic planetary models we expect the stable stratification would not extend all the way to the planetary radius but sit below a convective region, \citep[e.g.][]{Wahl2017,Debras2019}. It is therefore informative to consider the consequences of varying the radius of the stratified layer. We now consider a buoyancy frequency as described in equation~\ref {eq:N2_layer} with $\beta\neq1$. As our numerical model requires a finite size solid core for regularity of the solutions, we must consider a non-zero inner core size $\alpha r_0$, therefore, along with $\beta$, we explore how this parameter alters the results. 

Figure~\ref{fig:diss_ab_loglog} shows how the frequency-averaged dissipation varies with $\alpha$ and $\beta$, for $\bar{N}=1$, $\nu=\kappa=10^{-6}$. Figure~\ref{fig:diss_vol_int} shows the same parameter values but in this case we have calculated the corresponding volume of the stratified layer for each pair of $\alpha$ and $\beta$ values. The darker coloured lines show the dependence on $\alpha$ for a fixed $\beta=1.0$, and the lighter lines show the dependence on $\beta$ for fixed $\alpha = 0.1$.  When fitted, the $\beta$ dependence is roughly $\bar{D} \propto \beta^5$ (see Figure~\ref{fig:diss_ab_loglog}), which we note is the same as the dependence on $r_0$ found in the travelling approximation when considering $\beta=1$, see equation \ref{eq:D_TV}. It is clear from Figure~\ref{fig:diss_vol_int} that there is a strong volume effect, for which total dissipation decreases as the volume of the stratified layer decreases. However as these results found in Figure~\ref{fig:diss_vol_int} do not show a strictly linear relationship, there are additional dependencies on $\alpha$ and $\beta$ that are independent of the volume of the layer.

\begin{figure*}[t]
	\centering
	\subfigure[$D_{total}$]{\includegraphics[width=0.46\textwidth]{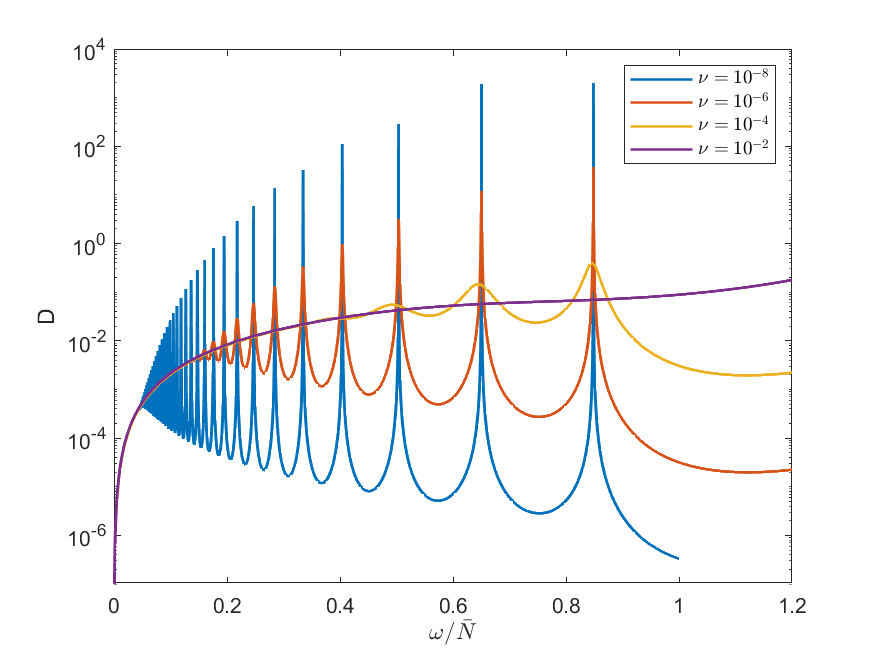}\label{fig:diss_nu}}
	\subfigure[$\omega=0.1$]{\includegraphics[width=0.46\textwidth]{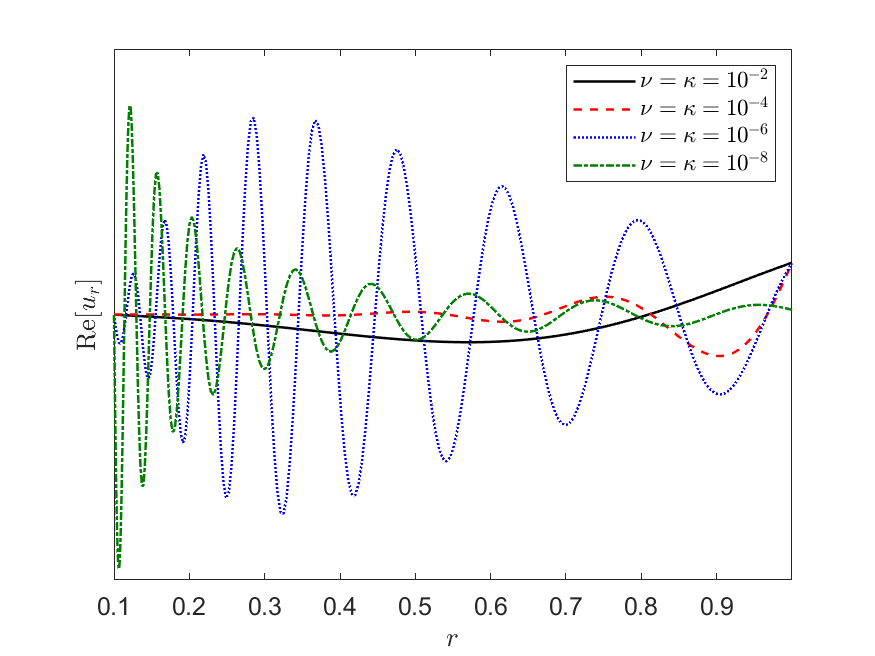}\label{fig:omega_nu}}
	\subfigure[Viscous dissipation]{\includegraphics[width=0.46\textwidth]{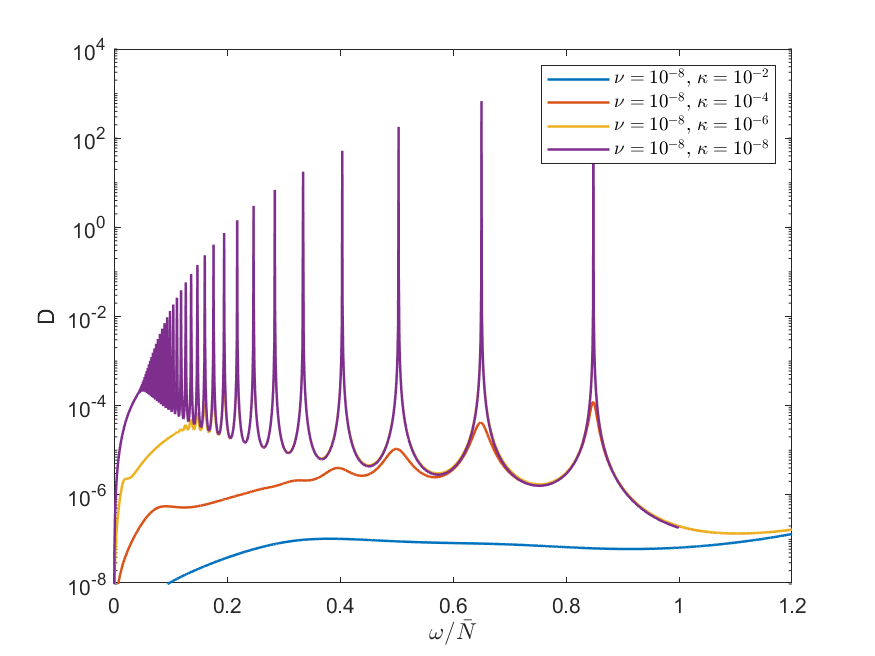}\label{fig:diss_Pr_visc}}
	\subfigure[Thermal dissipation]{\includegraphics[width=0.46\textwidth]{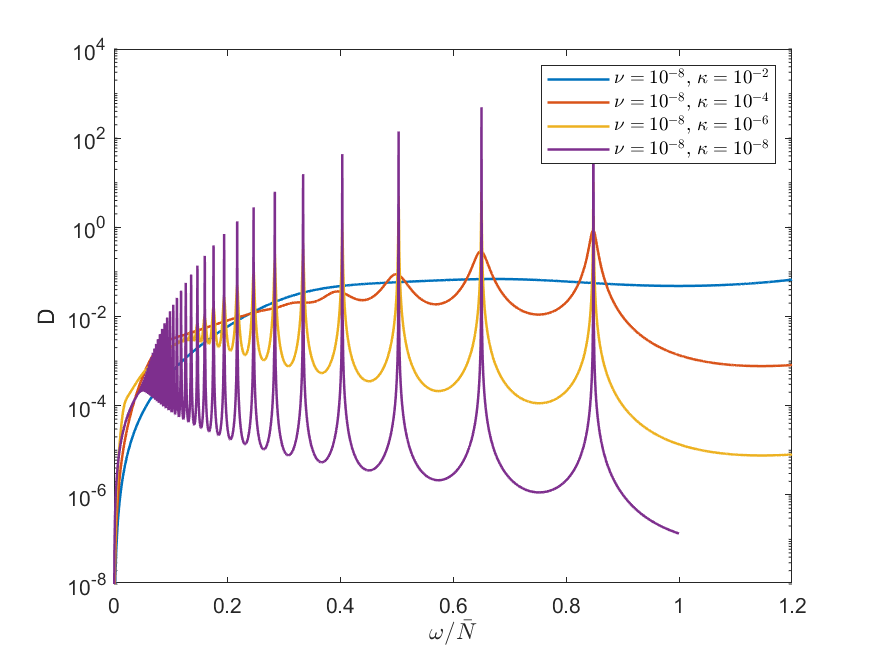}\label{fig:diss_Pr_ther}}
		\caption{Panel \protect\subref{fig:diss_nu} shows total dissipation as a function of frequency for different viscosities and thermal diffusivities with $Pr=1$. Panel \protect\subref{fig:omega_nu} shows the radial dependence of $u_r$ at forcing frequency of $\omega=0.1$ for the given parameters. Panels \protect\subref{fig:diss_Pr_visc} and \protect\subref{fig:diss_Pr_ther} show the viscous and thermal dissipation, respectively, for $\nu=10^{-8}$.}\label{fig:Pr_over}
\end{figure*}

This result is due to a combination of factors altering the amplitude of the forced response, which can be related to the rate of dissipation. Firstly, the waves are launched from the outer edge of the stratified layer at $\beta r_0$ and propagate inwards. The initial amplitude of the forced wave depends on this radius and decreases as $\beta$ decreases. However for standing waves the amplitude increases as the wave propagates inwards, leading to larger rates of dissipation close to the inner core. 

The frequencies of the internal gravity modes depend strongly on $\alpha$ and $\beta$. This is expected as the resonant frequencies depend on the half-integer wavelengths that fit inside the stratified layer, as shown by equation~\ref{eq:g_ana}. The solutions for a forcing frequency of $\omega=0.1$ are shown in Figure~\ref{fig:omega_ab}, where the differences in wavelength are clear. The green dashed line shows the case where the stratified layer only extends to half the planetary radius, where the wavelike behaviour can be seen to be confined to the stratified region and not persist in the convective region, in which the solution is evanescent. 

\subsubsection{Varying viscosity, thermal diffusivity and Prandtl number}\label{sec:param_visc}

The interiors of giant planets are, like most astrophysical fluids, expected to consist of regions of low viscosity, thermal diffusivity and Prandtl number. It is often numerically unrealistic to carry out studies in this parameter regime as the time and length scales required to be resolved are extremely small. For the purposes of our calculations, this means we would require very high radial resolution to resolve tidally forced waves with realistic planetary properties. Therefore, studies like ours must focus on parameter values that are more numerically achievable. In this section we also consider numerically convenient values, which allows us to explore the likely trends in dissipation rates. 

Initially we consider cases with a constant Prandtl number, while varying viscosity $\nu$, and thermal diffusivity $\kappa$, simultaneously. In Figure \ref{fig:diss_nu_int} for which $\alpha=0.1$, $\beta=1.0$, $\bar{N}=1$ we see the frequency-averaged dissipation has little dependence on viscosity and thermal diffusivity. This is a useful result on average for planetary applications, since realistic parameters are likely to be far from those considered here. This means that the overall frequency-averaged dissipative properties are insensitive to diffusivities, even if the dissipation at a given frequency is strongly dependent on these values.  

\begin{figure}
	\centering
	\subfigure[Fixed step width, $\delta r = 0.03$]{\includegraphics[width=0.45\textwidth]{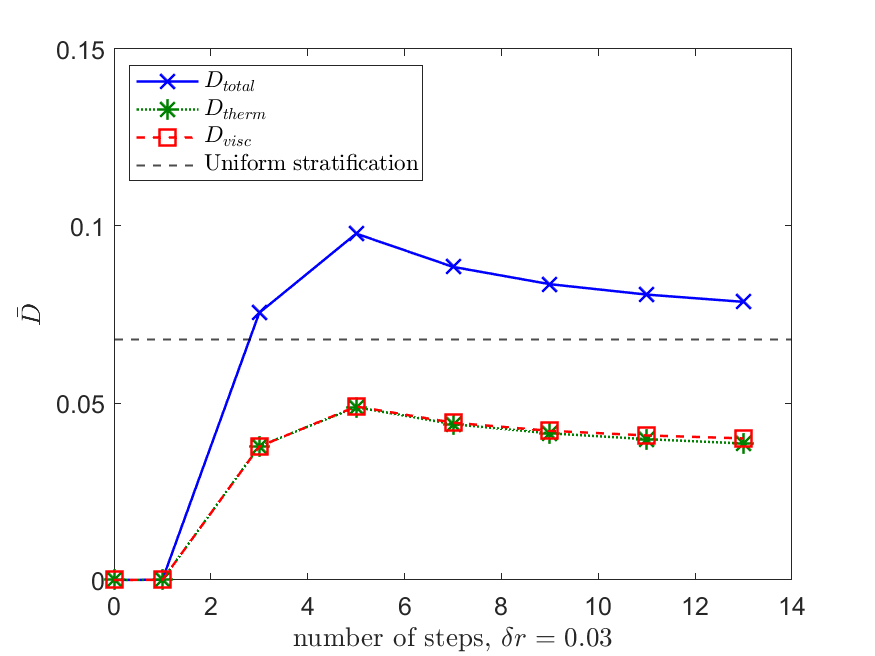}\label{fig:diss_step_delr}}
	\subfigure[Fixed total step width, $\sum \delta r = 0.54$]{\includegraphics[width=0.45\textwidth]{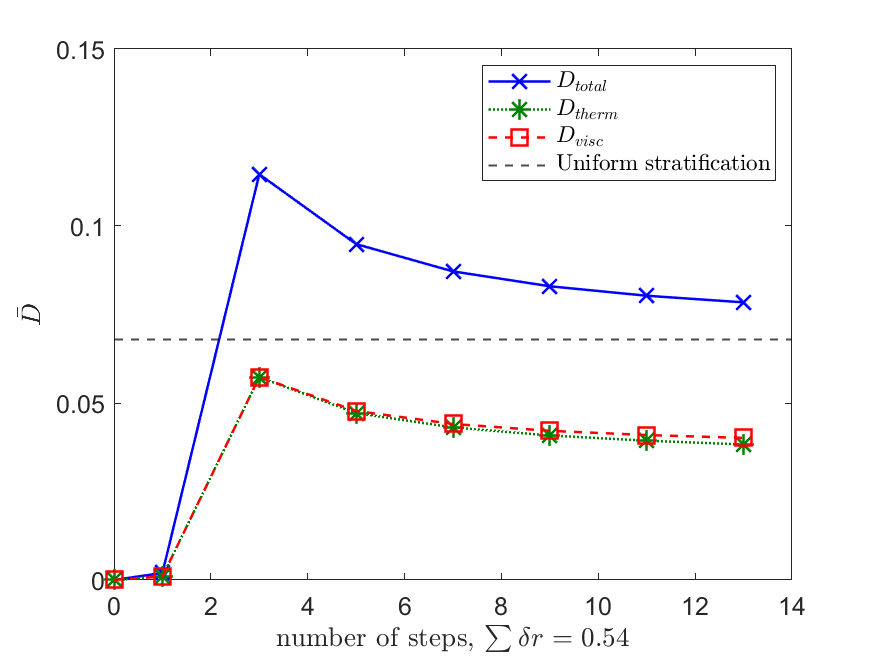}\label{fig:diss_step_sum}}
	\caption{Frequency-averaged dissipation as a function of the number of steps in a staircase, for $\alpha=0.1$, $\beta=1.0$, $\bar{N}=1$, $\nu=\kappa=10^{-6}$.}\label{fig:step_params}
\end{figure}

\begin{figure}
	\centering
	\subfigure[$\delta r= 0.03$]{\includegraphics[width=0.46\textwidth]{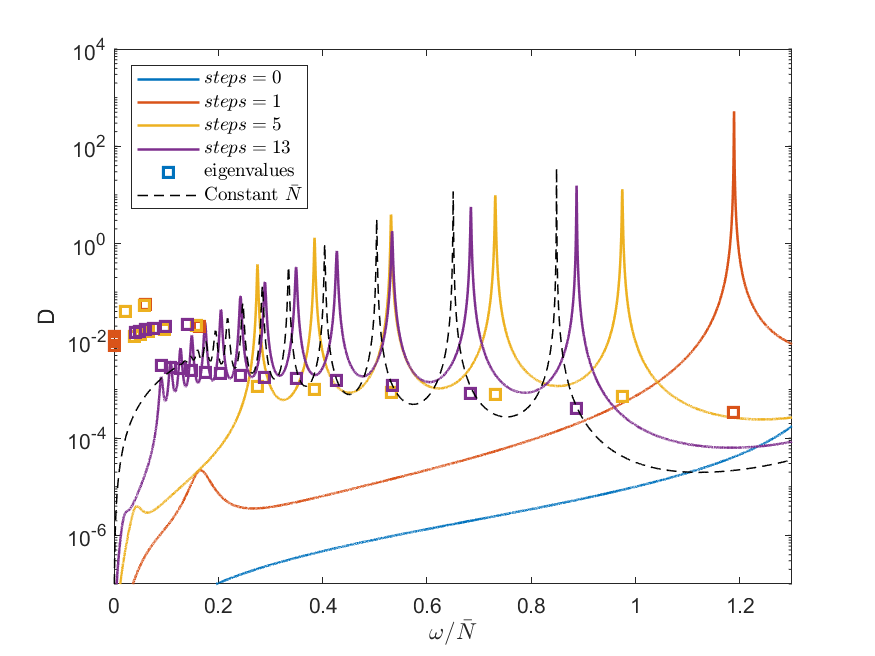}\label{fig:diss_step}}
	\subfigure[$\sum \delta r= 0.54$]{\includegraphics[width=0.46\textwidth]{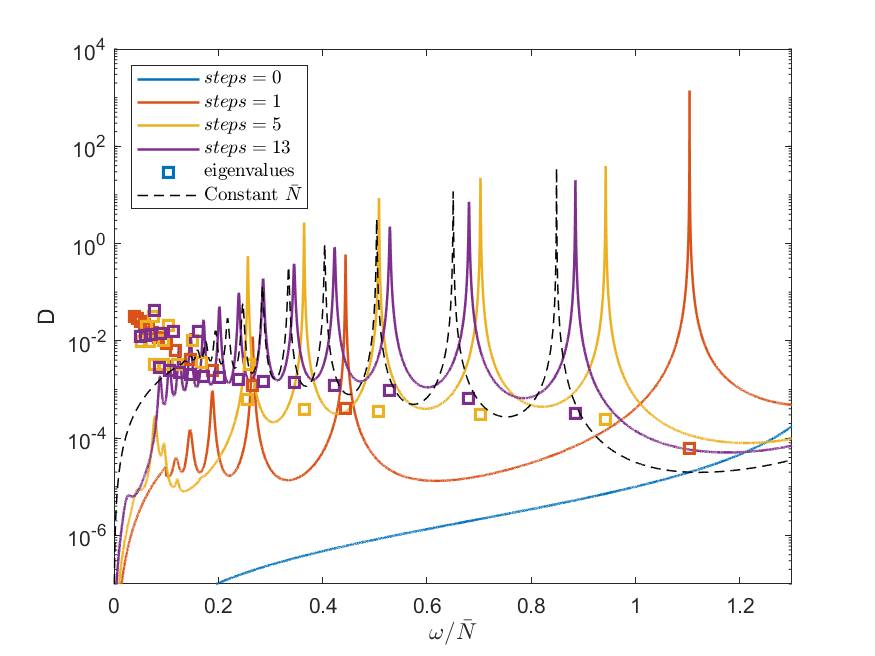}\label{fig:diss_step_2}}
		\caption{Dissipation as a function of frequency for different numbers of steps with fixed $\alpha=0.1$, $\beta=1.0$, $\bar{N}=1$, $\nu=\kappa=10^{-6}$. Panel \protect\subref{fig:diss_step} shows the case for fixed step width $\delta r=0.03$. Panel \protect\subref{fig:diss_step_2} shows the case for fixed total step width $\sum \delta r =0.54$. } \label{fig:step_over}
\end{figure}

This can be explained by considering the low-frequency and mid-frequency regimes separately. For frequencies sufficiently low that the travelling wave approximation is valid (see \S~\ref{sec:trav_wave}), the result is found to be independent of damping mechanism and therefore each of $\nu$, $\kappa$ and $Pr$. To consider the mid-frequency range where resonant peaks are clear, we refer back to the analogy of a forced damped harmonic oscillator discussed in \S~\ref{sec:SHO}. There we established that the width of the peak was proportional to the damping rate, and the height of the peak inversely proportional to the damping rate. Therefore, upon integration, they have counteracting effects on the frequency-averaged dissipation; this is confirmed in Figure~\ref{fig:diss_nu}, where these effects are visible. Although the point at which we transition into the travelling wave regime varies considerably, this has no effect on the overall results. 

In Figure \ref{fig:diss_nu} we also observe the effects of viscosity on internal gravity mode frequencies is very small, being unobservable on these plots. Instead, the mode frequency is strongly dependent on the location and strength of the stable stratification, which is kept constant across these cases. 

Figure~\ref{fig:omega_nu} shows the radial dependence of $u_r$ at a tidal forcing frequency of $\omega=0.1$ for different viscosities and thermal diffusivities with a fixed Prandtl number $Pr=1$. At this tidal forcing frequency we are below the regime in which a discrete set of resonances are visible at all but the lowest viscosity, and it is clear in the case of the largest viscosity that we are in the travelling wave regime. We observe that, although the integrated dissipation is independent of viscosity, the spatial structure of the response, like the frequency-dependent dissipation, is sensitive to viscosity and thermal diffusivity.

\begin{figure}
	\centering
	\subfigure[Varying $\bar{N}$ with fixed $\nu=\kappa=10^{-6}$, steps$=3$]{\includegraphics[width=0.45\textwidth]{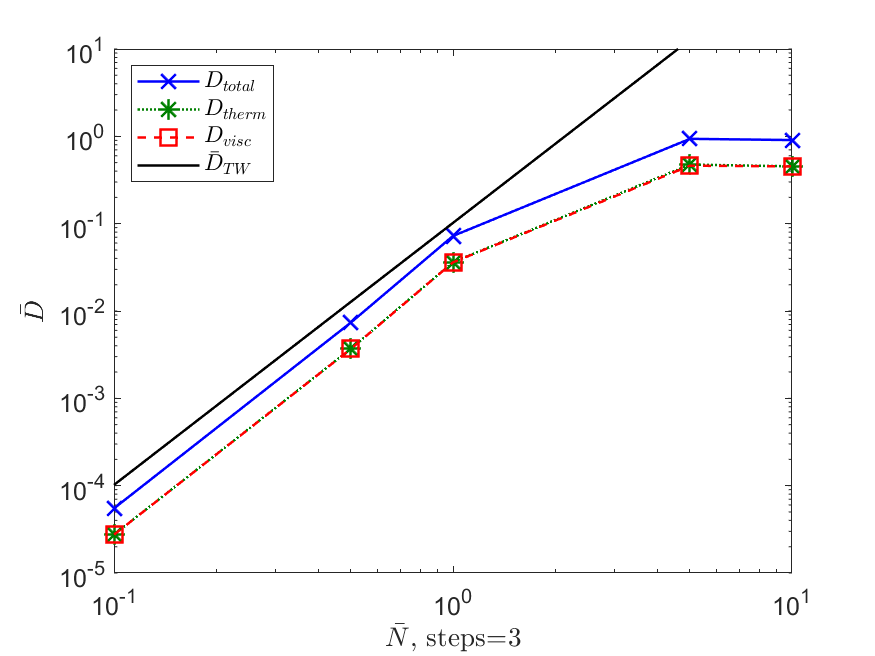}\label{fig:diss_step_N}}
	\subfigure[Varying $\bar{N}$ with fixed $\nu=\kappa=10^{-6}$, steps$=3$]{\includegraphics[width=0.45\textwidth]{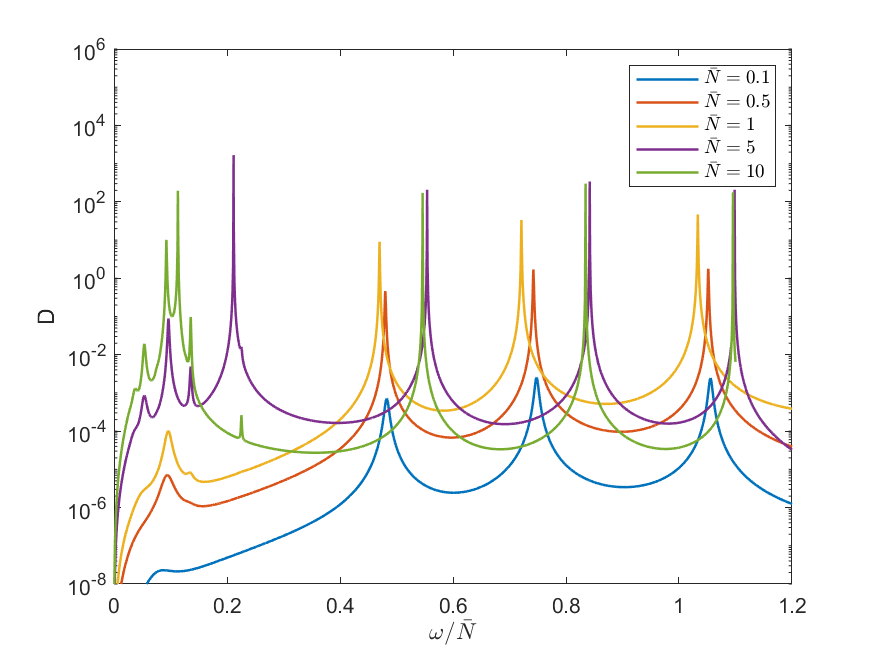}\label{fig:omega_N_step}}
	\caption{Panel~\protect\subref{fig:diss_step_N} shows the frequency-averaged dissipation as a function of stratification strength, $\bar{N}$ for the case of a staircase-like density structure with other parameters fixed at $\alpha=0.1$, $\beta=1.0$, steps $=3$, $\delta r=0.06$, $\nu=\kappa=10^{-6}$. Panel~\protect\subref{fig:omega_N_step} shows the corresponding frequency dependence.} \label{fig:step_N}
\end{figure}

Turning our attention now to varying the Prandtl number, Figure \ref{fig:diss_Pr_nu_int}  and \ref{fig:diss_Pr_kappa_int} show the frequency-averaged dissipation as a function of Prandtl number. We consider two cases, one of fixed viscosity $\nu$ = $10^{-8}$, and the other fixed thermal diffusivity $\kappa$ = $10^{-2}$; both cases again fix $\alpha=0.1$, $\beta=1.0$ and $\bar{N}=1$ as before. We find the integrated dissipation $D_{total}$ has very little dependence on Prandtl number and the main consequence of decreasing Prandtl number is the subsequent decrease in the ratio of $D_{visc}$ to $D_{ther}$; we also note the similarity between the figures despite the significant variation in total diffusivity $\nu+\kappa$. Looking at the frequency dependence in Figures \ref{fig:diss_Pr_visc} and \ref{fig:diss_Pr_ther}, we see the width and height of the resonant peaks also exhibit the behaviour of the forced damped harmonic oscillator. 

\subsection{Parameter dependences for semi-convective layers}\label{sec:param_step}

As it is possible that stratified layers in giant planets could be unstable to double-diffusive convection, leading to the presence of staircase-like density structures, we now consider how semi-convective layers can alter the measured dissipation. Instead of a uniformly stratified medium, we now consider a series of convective steps with stably stratified interfaces, as described by equations \ref{eq:N2_steps}, which have an equivalent mean stratification to a uniformly stratified layer of equal depth.

\subsubsection{Convergence with step number}

Shown in Figures \ref{fig:diss_step_delr} and \ref{fig:diss_step_sum} is the frequency-averaged dissipation as a function of step number for cases in which the staircase extends across the entire depth of the planet. In both cases we fix $\alpha=0.1$, $\beta=1$, $\bar{N}=1$ and $\nu=\kappa=10^{-6}$, but in Figure \ref{fig:diss_step_delr} the width of each interface is fixed, and in Figure \ref{fig:diss_step_sum} the total width of all interfaces is fixed with the height of the peaks adjusted accordingly. Although it is likely that a staircase structure does not extend all the way to the planetary surface, we have considered $\beta=1$ in these cases to reduce the required radial resolution, allowing us to consider cases with a higher step number whilst keeping numerical demand low. In both cases the total dissipation $\bar{D}_{total}$ for the equivalent uniformly stratified case has been plotted (black-dashed line).

Although for very low step numbers these cases exhibit slightly different trends, where the total dissipation is larger than that of a uniformly stratified layer, the total frequency-averaged dissipation very quickly converges to that of the uniform stratified layer. The initial rapid increase of the dissipation as we go from 1 to 3 steps is presumably related to the outer interface moving closer and closer to the surface with increasing step number (where the non-wavelike tidal response in the outermost convective region is larger in amplitude leading to more efficient wave excitation). There is large uncertainty in the number of steps that could form, and it is likely to evolve with time as they merge \citep{Wood2013,Belyaev2015}; however this suggests that unless there are very few layers the staircase will act similarly to a uniformly stratified medium on average (in our non-rotating model), although the details will depend on the frequency of the tidal forcing. 

The frequency dependence of the dissipation for different step numbers is shown in Figures~\ref{fig:diss_step} and~\ref{fig:diss_step_2} for fixed step width and fixed total step width respectively. We see additional peaks arising with increasing step number and observe internal gravity mode resonances at low frequencies. These short wavelength modes have formed within the finite widths of the interfaces; with fixed step size they occur for $\omega<0.2$, and have little contribution to the frequency-averaged dissipation. However, with fixed total step width they occur at much higher frequencies ($\omega<0.5$ in steps $=1$ case). This is due to the large interface width in the case of a single layer, allowing for internal gravity waves with comparably large wavelengths to form.

\subsection{Varying the strength of the staircase stratification}

Similarly to the uniformly stratified case, we now consider how the frequency-averaged dissipation depends on the strength of the stratification in Figure \ref{fig:diss_step_N}. We see the trend found in \S~\ref{sec:param_N}, and the agreement with the travelling wave calculation (shown here as $\bar{D}_{TW}$) holds here also as the staircase-like structure behaves like a uniform medium on average. This suggests that results discussed there are also relevant here, and looking at the dissipation profile in Figure \ref{fig:omega_N_step}, we see the same narrowing of peaks as stratification strength increases. The notable difference between these cases is the behaviour of the g-modes within the interfaces. Resonances with these modes increase significantly with stratification, and in fact even with only moderate stratification ($\bar{N}=5$) they have magnitudes comparable to the interfacial modes, and therefore contribute significantly to the frequency-averaged dissipation. 

\begin{figure}
	\subfigure[$\alpha=0.1$, $\bar{N}=1$, $\delta r=0.1$, $\nu=\kappa=10^{-6}$]{\includegraphics[width=0.46\textwidth]{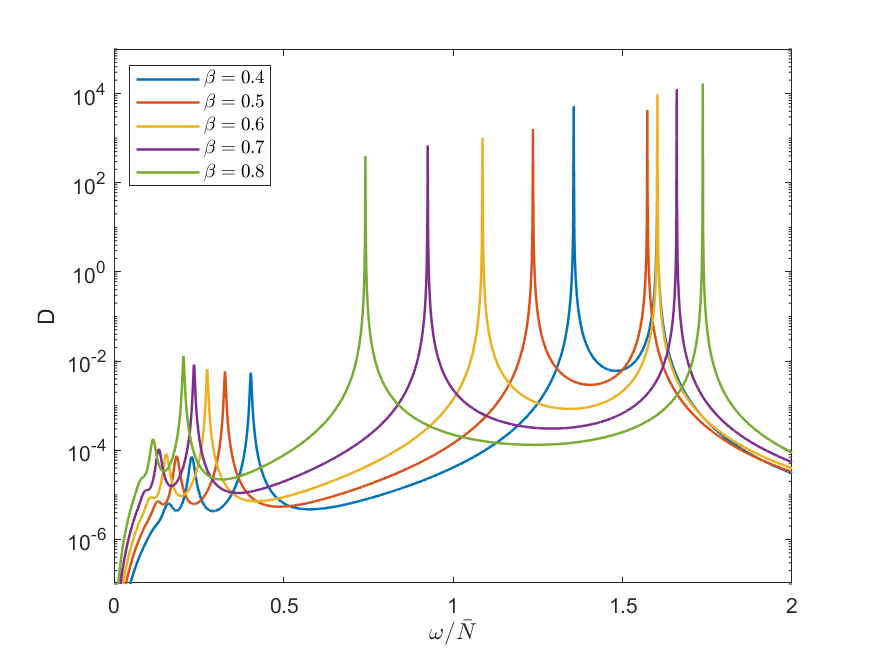}\label{fig:diss_He}}
	\subfigure[$\alpha=0.1$, $\bar{N}=1$, $\delta r=0.1$, $\nu=\kappa=10^{-6}$, $\beta=0.7$]{\includegraphics[width=0.46\textwidth]{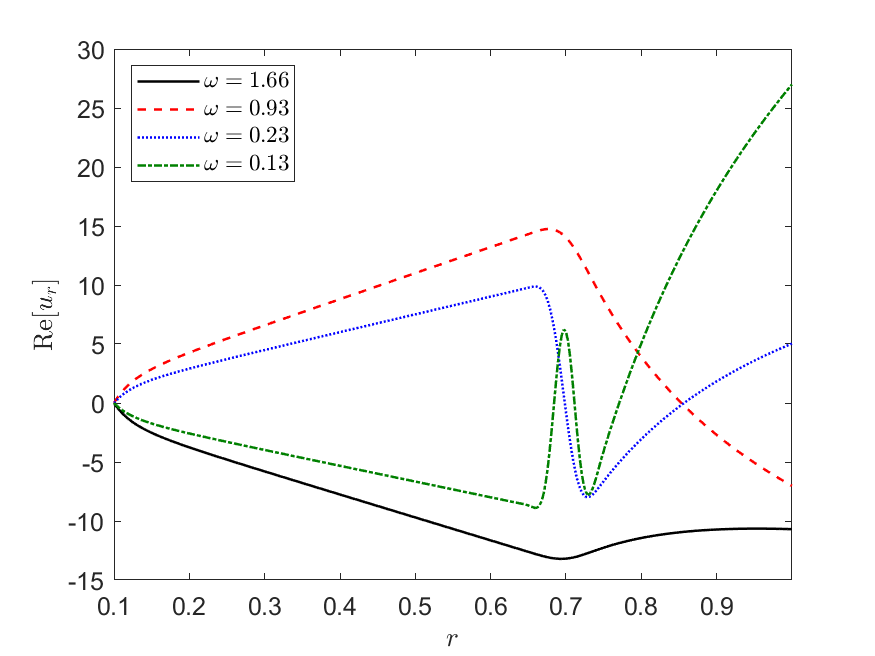}\label{fig:omega_He}}
	\caption{Panel  \protect\subref{fig:diss_He} shows dissipation as a function of frequency for an isolated stable layer at different radii. Panel  \protect\subref{fig:omega_He} shows the corresponding radial dependence of the forced solution at different frequencies. } \label{fig:He_over}
\end{figure}

\begin{figure}
	\centering
	\subfigure[$\omega_{min}=0$, $\omega_{max}=2$]{\includegraphics[width=0.45\textwidth]{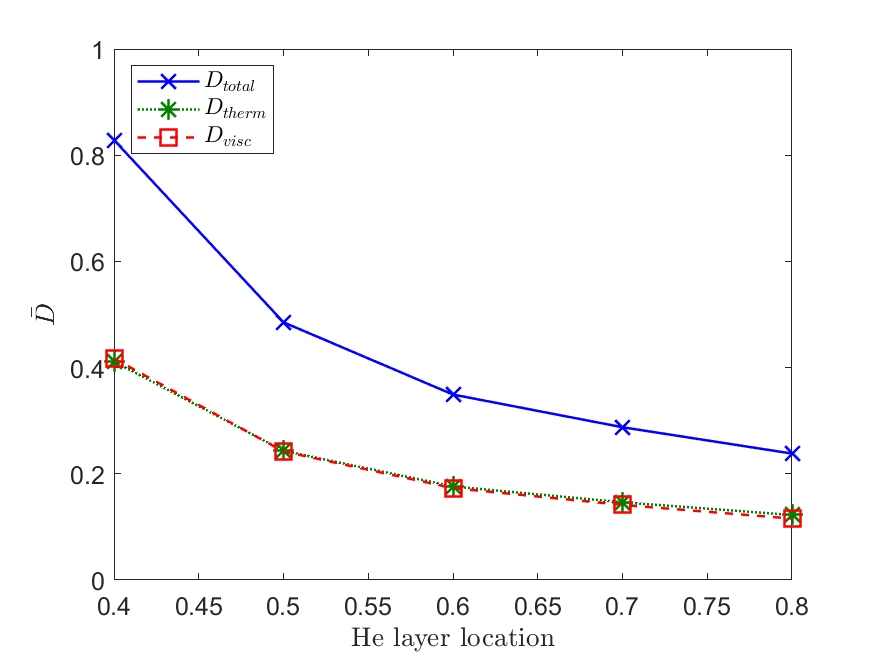}\label{fig:diss_He_0_2}}
	\subfigure[$\omega_{min}=0$, $\omega_{max}=0.6$]{\includegraphics[width=0.45\textwidth]{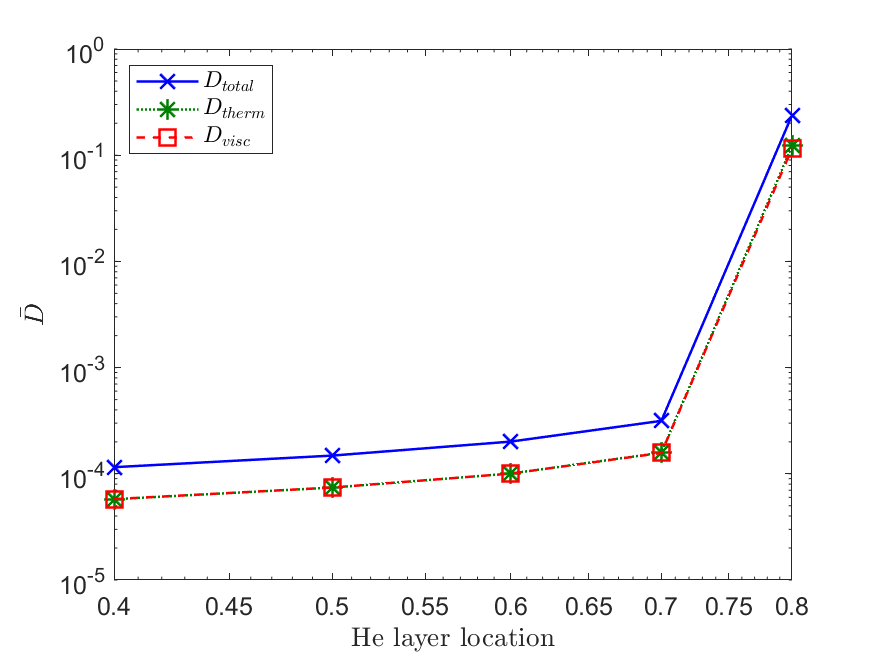}\label{fig:diss_He_0_06}}
	\caption{Frequency-averaged dissipation as a function of helium layer location in radius, for cases with $\alpha=0.1$, $\bar{N}=1$, $\delta r=0.1$, $\nu=\kappa=10^{-6}$.} \label{fig:He_params}
\end{figure}

\subsection{Isolated stable layer - helium rain layer}
 
Having explored a stable layer just outside the core, we now consider an isolated stable layer near the transition region between the metallic and molecular regions, potentially caused by helium rain \citep[e.g.][]{Stevenson1977,Nettelmann2015}. We use equation~\ref{eq:N2_steps} to describe a single wide step ($\delta r=0.1$) to represent an isolated transition layer. In Figure~\ref {fig:diss_He} we show total dissipation ($D_{total}$) for cases where $\bar{N}=1$, $\nu=\kappa=10^{-6}$, and $\alpha=0.1$. The location of the wide interface is varied using $\beta$, which here represents the centre of the layer. We see the large surface gravity resonance about $\omega=1.6$, the single interfacial mode (where the layer oscillates as an interface) between $\omega=0.6$ and $\omega=1.5$, and a series of shorter wavelength internal gravity mode resonances at low frequencies. 

There are two things to note in this figure. First, the frequency of the interfacial mode depends strongly on the location of the layer $\beta$, which could have consequences when considering observations of mode-mixing or the possibility of resonance-locking \citep{Fuller2014,Fuller2016,Dewberry2021}. Second, the amplitude of the interfacial and surface gravity mode peaks decrease in height as the layer moves outwards whilst the internal gravity waves peaks increase in magnitude. Figure \ref{fig:He_params} shows the frequency-averaged dissipation, and we can see in panel \protect\subref{fig:diss_He_0_2} where the upper frequency limit has been taken to be $\omega_{max}=2$, that dissipation decreases as the layer is moved outwards. However, in panel \protect\subref{fig:diss_He_0_06} we take $\omega_{max}=0.6$ so that the integrated quantity only includes the internal gravity wave resonances within the step; there is an increase in dissipation. Therefore, the importance and consequences of an isolated stably stratified layer depends on whether the forcing frequency is close to the interfacial mode resonance or the internal gravity mode resonances. 

\begin{figure}[h]
	\centering
\subfigure{\includegraphics[width=0.46\textwidth]{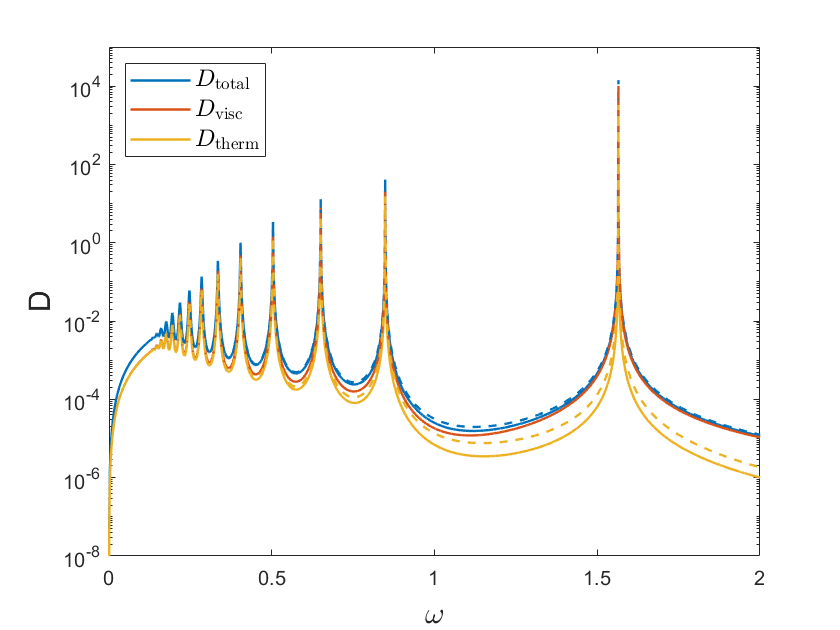}}\label{fig:diss_altg}
	\caption{Dissipation rates for a case considering the gravity profile of a dense core. Parameter values are $\alpha=0.1$, $\beta=1.0$, $\bar{N}=1$, $\nu=\kappa=10^{-6}$. The solid lines show results for $g \propto \frac{1}{r^2}$ and the dashed lines $g \propto r$ for comparison.} \label{fig:altg_over}
\end{figure}

\subsection{Gravity profile for a dense core with $g \propto 1/r^2 $}\label{sec:alt_g}

So far we have considered the gravity profile of a homogeneous body where $g \propto r$, which is usually adopted in the Boussinesq approximation. We do not expect a planet to be homogeneous and we therefore consider how changing this profile to the gravity field of a centrally condensed body affects our results. To do this, we consider $g \propto \frac{1}{r^2}$, which results in the following governing equations, but we note that the surface gravity $g_0$ is still 1,
\begin{equation}\label{eq:mtm_altg}
\frac{\partial \vb{u}}{\partial t} = - \frac{1}{\rho_0} \nabla p + \frac{b}{r^2} \hat{r} + \nu \nabla^2 \vb{u} - \nabla \psi,
\end{equation}
\begin{equation}\label{eq:b_altg} 
\frac{\partial b}{\partial t} + r^2 u_r N^2 = \kappa \nabla^2 b,
\end{equation}
with similar changes in the boundary conditions and dissipation rates. The buoyancy force and gravity term appear on their own, in the same term in equation~\ref{eq:mtm_altg}, and in the absence of thermal diffusivity, the terms in equation~\ref{eq:b_altg} can be represented similarly. Hence, we obtain identical analytical predictions for the free mode frequencies to the homogeneous case, and differences in the forced response arise solely in the buoyancy equation directly affecting thermal dissipation only. Hence, we expect the dissipative responses to be very similar for both gravity profiles.

Our numerical results confirm this prediction, and we find that changing the gravity profile does not significantly alter our conclusions. As expected the resulting velocity, and viscous dissipation, remain unchanged and only the thermal dissipation varies. Figure~\ref{fig:altg_over} shows the total, viscous and thermal dissipation rates with the gravity profile of a dense core (solid line) compared with that of a homogeneous body (dashed line), where we note only minor quantitative differences. 

\section{Comparison to realistic values and observed tidal quality factors}\label{sec4}

\begin{figure}
	\centering
	\subfigure{\includegraphics[width=0.46\textwidth]{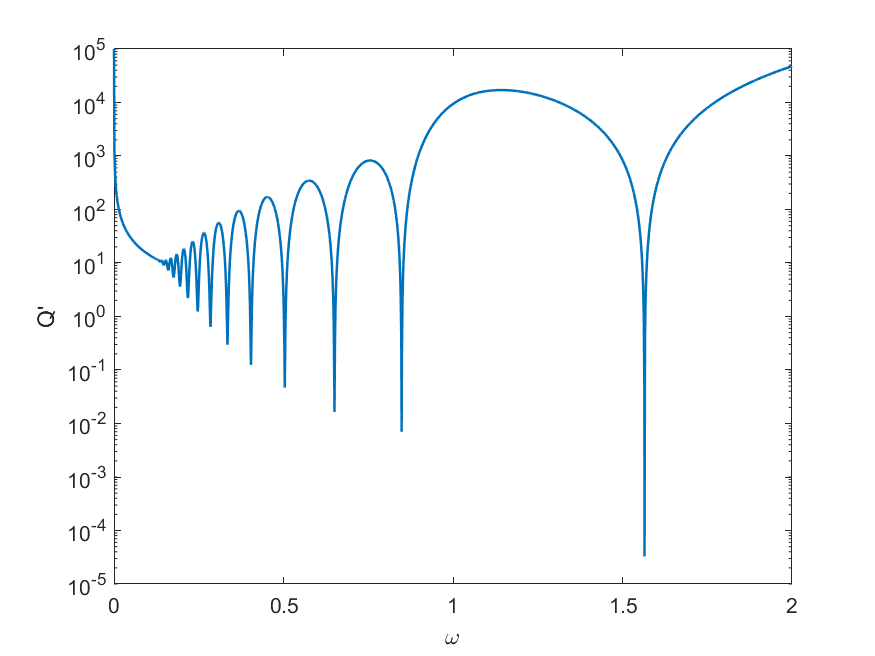}}
	\subfigure{\includegraphics[width=0.46\textwidth]{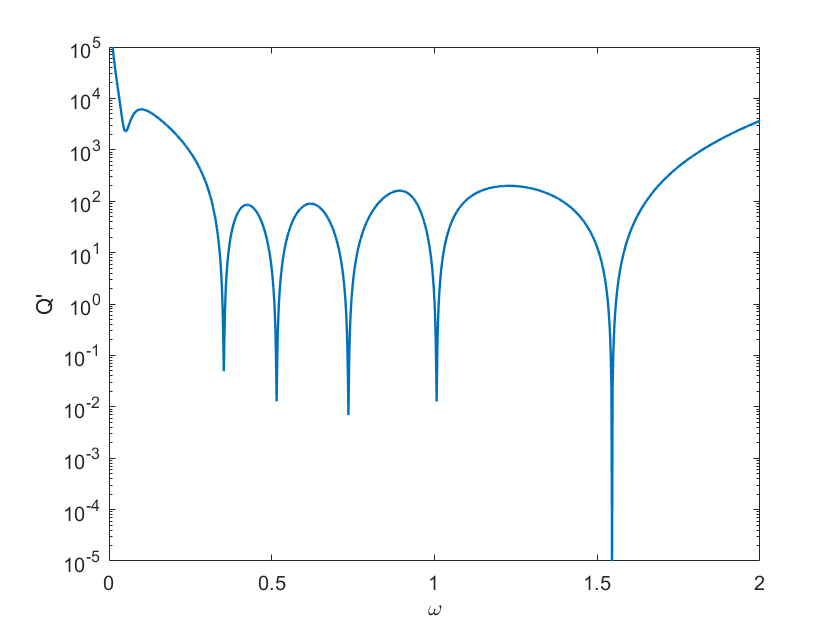}}
	\caption{Top: example tidal quality factor profile for the case with uniform stratification shown in Figure~\ref{fig:diss_over_a} with $\alpha=0.1$, $\beta=1.0$, $\bar{N}=1$, $\nu=\kappa=10^{-6}$. Bottom: same for a staircase with 5 steps with $\delta r=0.03$ as shown in Figure~\ref{fig:diss_step}, with $\alpha=0.1$, $\beta=1.0$, $\bar{N}=1$, $\nu=\kappa=10^{-6}$.} \label{fig:tidalQ}
\end{figure}

We have discussed a highly simplified model, and values for tidal dissipation rates should not be taken too far out of this context. However, it is still informative to consider the extrapolation of our results to compute tidal quality factors to appreciate the astrophysical significance of our results. Following \cite{Ogilvie2014}, the total tidal dissipation rate can be related to the dimensionless complex Love number by
\begin{equation}
D=|\omega|\frac{(2l+1) r_0 |\mathcal{A}|^2}{8 \pi G} \operatorname{Im}[k^m_l(\omega)].
\end{equation} 
Therefore considering for a homogeneous fluid body that $\operatorname{Im}[k^m_2(\omega)]=\frac{3}{2 Q'}$ and  $\mathcal{A}=\psi_0 r_0^2$, we can write the modified tidal quality factor, 
\begin{equation}
Q'=|\omega| \frac{(2l+1) r_0^3 |\psi_0|^2}{8 \pi G} \frac{3}{2 D}.
\end{equation} 
Therefore, using the non-dimensionalisation used previously, noting that $G=1$ in our units, the tidal quality factor can be written in terms of our variables as,
\begin{equation}
Q'=|\omega| \frac{(2l+1) |\psi_0|^2}{8 \pi } \frac{3}{2 D}.
\end{equation} 
Since we consider a linear problem, $D \propto \psi_0^2$ and therefore $Q'$ is independent of tidal amplitude. Setting $l=2$ and $|\psi_0|=1$ as we have considered in our numerical results, we find
\begin{equation}
Q'=\frac{15 |\omega|}{16 \pi D}.
\end{equation}
Therefore, taking our first example of a dissipation profile shown in Figure~\ref{fig:diss_over_a}, we can find the frequency-dependent tidal quality factor shown in Figure~\ref{fig:tidalQ}. 

In the travelling wave regime equation (\ref{eq:D_TV}) gives
\begin{equation}\label{eq:Q_convert}
Q'=\frac{30 \sqrt{6} }{16 \pi}\frac{\omega_d^2}{\bar{N} |\omega|}.
\end{equation}
From this it is possible to calculate quantities such as tidal circularisation timescales. From \cite{Barker2020}, the timescale for circularisation due to dissipation in a planet with mass $M$ due to a body with mass $M_2$ is
\begin{equation}
\tau_{e}=\frac{2}{63\pi}Q' \bigg(\frac{M}{M_2}\bigg)\bigg(\frac{M+M_2}{M}\bigg)^{\frac{5}{3}}\frac{P_{orb}^{\frac{13}{3}}}{P_{dyn}^{\frac{10}{3}}}.
\end{equation}
We adopt some approximate numbers for illustration for the hot Jupiter WASP-12b, using data taken from \cite{EUExo}, $M=1.5M_J=2.8\times 10^{27}$kg, $M_2=1.4 M_{\odot}=2.9\times10^{30}$kg, $P_{orb}=2\pi/\Omega_o=1.1$ days, $r_0=1.9R_J$, $P_{dyn}=2\pi/\omega_d=0.26$ days. For eccentricity tides $\omega=\Omega_o=0.235\omega_d$, and at this tidal frequency, Figure~\ref{fig:tidalQ} gives $Q'\sim 10$, and equation \ref{eq:Q_convert} predicts a similar value if we assume the travelling wave regime. Thus, we obtain
\begin{equation}
\tau_{e}\approx 3.92~\text{years},
\end{equation}
which shows very efficient tidal circularisation is possible with the excitation of internal waves if the planet is stably stratified throughout a significant portion of its radius with $\bar{N}\sim\omega_d$. If instead, only a fraction $\beta\sim0.5$ of the planet is stably stratified, we would expect the timescale to be larger by a factor of $\beta^5 \sim 32$ (based on \S~\ref{sec:param_size}), which would still predict very efficient dissipation for WASP-12b. In this case the travelling wave regime is highly likely to be initiated by nonlinear effects such as wave breaking \citep[e.g.][]{Barker2010} rather than thermal diffusion or microscopic viscosity, as dimensionless tidal amplitudes are very large in WASP-12b (we have $\frac{m_2}{M}\left(\frac{r_0}{a}\right)^3\sim 0.05$, so the waves are highly likely to be nonlinear). Note that the dissipation rate in the travelling wave regime is insensitive to the damping mechanism (and microscopic diffusivities) as long as the waves are efficiently damped.

We would also predict smaller planets such as hot Neptunes to circularise very rapidly due to stable layers with a similar calculation: using $M=10^{26}$kg, $r_0=2.46\times 10^7$m, $M=2\times 10^{30}$kg, $P_{orb}=5$ days, gives $\tau_e\approx 3.6\; \mathrm{Myr}$ if $Q'\sim 100$, as would be predicted by  equation \ref{eq:Q_convert} or Figure~\ref{fig:tidalQ} for tidal frequencies $\omega\approx 0.022\omega_d$ if $\bar{N}\sim \omega_d$.

Similarly, following e.g.~\cite{Barker2020} for the timescale of outward orbital migration of a moon from an initial orbit with semi-major axis $a$ to an orbit with semi-major axis $2a$ due to dissipation inside the planet, we find
\begin{equation}
\tau_{a}=\frac{4}{117} \big(2^{\frac{13}{2}} -1 \big) \frac{Q'}{\Omega_o} \bigg(\frac{M}{M_2}\bigg)\bigg(\frac{a}{r_0}\bigg)^{5},
\end{equation}
where we have assumed $\Omega_0 \propto a^{-\frac{3}{2}}$ and constant $Q'$.
Using values for Io and Jupiter taken from the \cite{JPL} $M=M_J=1.90 \times 10^{27}$kg, $M_2=8.93\times10^{22}$kg, $r_0=R_J=6.99\times 10^4$km, $a=4.22\times 10^5$km, and assuming a rotation period $2\pi/\Omega=10$ hrs, we find a tidal frequency $\omega= 2(\Omega-\Omega_o)\approx0.44\omega_d$. For such values Figure~\ref{fig:tidalQ} gives $Q'\in [10,10^2]$, and equation \ref{eq:Q_convert} would predict $Q'\sim 3$ (if the travelling wave regime applies). If we assume $Q'=10$ we find
\begin{equation}
\tau_{a}\approx 4\; \mathrm{Myr}.
\end{equation}
If applicable, this mechanism is therefore sufficient to have played a major role in the evolution of the Galilean moons \citep[e.g.][]{Lainey2009,Lainey2017}. Although such efficient dissipation rates would require large portions of the planet to be stably stratified, this calculation indicates the potentially huge importance of such layers on tidal dissipation rates. On the other hand, if the travelling wave regime does not apply, then it should be noted that in linear theory much weaker dissipation rates would be predicted in between the discrete resonances if we were to adopt realistic (computationally inaccessible) values of $\nu$ and $\kappa$. We will consider a more realistic model incorporating rotation in a subsequent paper \cite[following][]{Pontin2022}.


\section{Conclusions}\label{sec:conc} 

We have used a combination of analytical and numerical approaches to explore how stably stratified layers affect tidal dissipation rates in planetary interiors. Our numerical calculations consider dissipation of internal waves through viscous forces and thermal damping excited by realistic tidal forcing, but we have adopted the Boussinesq approximation to allow more detailed understanding. We represent uniform stably stratified layers that extend to a specified radius, or a staircase-like density structure that might be produced by double-diffusive convection, by a numerically smooth profile.

As expected from previous studies \citep[e.g.][]{Ogilvie2004,Ogilvie2009,Fuller2016,Andre2019}, the dissipation rate is observed to be strongly frequency dependent with clear resonances with the free modes of the system. In this non-rotating case, when considering uniform stratification we observe internal gravity modes, resonant within the width of the layer. For a staircase-like density profile, we observe interfacial modes as well as short wavelength internal gravity modes that form within the width of the interfaces. In both cases we observe a large resonance with the surface gravity mode at the planetary surface. We found good agreement with both the oscillation frequency and damping rate with the numerical eigenvalue problem and with analytical calculations of the free modes, consistent with studies of the non-dissipative system \citep[e.g.][]{Andre2017,Pontin2020}. 

We quantify the dissipation using a frequency-averaged quantity as a way to determine the overarching trends as the parameters are varied. Using this quantity, we establish that the dissipation rate is strongly dependent on the stratification strength $\bar{N}$, and depends on radius of stratification as $\propto \beta^5$. These results agree with the corresponding analytical calculations in the travelling wave regime. We found the integrated dissipation rates to be largely independent of viscosity and thermal diffusivity, despite the differences in the frequency-dependent profiles. This is a useful result as our parameter values necessarily lie outside the range of realistic values due to computational constraints. This is also in agreement with the analogy with a damped harmonic oscillator for higher frequencies, where we established that the peak width is proportional to the damping rate, and peak heights are inversely proportional to damping, resulting in counteracting effects in the integrated quantity. 

When comparing a staircase-like profile with an equivalent uniformly stratified layer, we found that provided there are a sufficient number of steps (a few) the staircase acts like that of a uniformly stratified layer on average. However, the presence of a staircase does alter the frequency-dependent dissipation, particularly in frequency ranges containing resonances, so it could impact the dissipation rate significantly depending on the tidal forcing frequency. When considering the case of an isolated stratified layer near the molecular/metallic transition region, we observe a large resonance corresponding to a single interfacial mode. This can have a significant contribution to the overall dissipation rate, and due to the magnitude of this single resonance, could have an impact for resonance-locking scenarios \citep{Fuller2016}. The stable layer also supports short wavelength internal gravity waves leading to low frequency resonances that contribute to the dissipation rate. 
 
Overall, we found that stable stratification, whether a uniform layer, a staircase-like structure, or a helium rain layer, can contribute to the dissipation rate by the introduction of additional resonances into the system. Even in cases where the contribution to the integrated dissipation quantity is low, the properties of the stable stratification can have a strong effect at particular frequencies, which if sufficiently close to those of the tidal forcing, could have important implications for tidal orbital and spin evolution \citep[similarly to the case of tides in radiation zones of stars, e.g.,][]{Goodman1998,Terquem1998,Barker2010}. We have crudely estimated the resulting tidal evolutionary timescales due to dissipation in such extended stable layers, finding that this mechanism could provide a crucial contribution to tidal dissipation rates in solar and extrasolar planets, including those that range from the masses of Neptune to Jupiter, which should be studied further.

\subsection{Future outlook}

This paper has focused on a very simplified non-rotating system to reduce numerical demand and allow for comprehensive analytical and numerical analysis of internal waves. However, Coriolis effects will have implications for tidal dissipation in giant planets, through the inclusion of inertial and gravito-inertial waves \citep[e.g.][]{Dintrans1999,Rieutord2009, Andre2019}. Previous studies show that tidal dissipation in convection zones primarily occurs through the excitation and dissipation of inertial waves \citep{Ogilvie2004,Wu2005a,Wu2005b,Favier2014}. We will present an analysis incorporating rotation in a follow-up paper \citep[building upon][]{Pontin2022}, and we note the wider community has also begun to explore this problem recently \citep[e.g.][]{Lin2023,Dewberry2023}. It would also be of interest to separate the thermal and compositional contributions to the buoyancy and study whether double-diffusive effects could be important in the tidal problem (beyond motivating our density profiles).
\\\\
\noindent
CMP was supported by an STFC PhD studentship 2024753.  
AJB was supported by STFC grants ST/R00059X/1, ST/S000275/1 and ST/W000873/1. RH was supported by STFC grants ST/S000275/1 and ST/W000873/1. We would like to thank the two referees for their positive and constructive reports that helped us to improve the manuscript. We would also like to thank St\'{e}phane Mathis and Quentin Andr\'{e} for discussions at an early stage in this project, and Chris Jones and Gordon Ogilvie for helpful feedback.

\appendix
\twocolumngrid
\section{Viscous dissipation}\label{app:dvisc}

It can be informative to separate the viscous dissipation into two components: the viscous dissipation occurring in the bulk of the fluid and the normal viscous flux through the boundary. Considering the momentum equation stated in equation \ref{eq:mtm} and using index notation, we write, 
\begin{equation}
\partial_t u_{i} = b g_i + \partial_j\sigma_{i j},
\end{equation}
where $ \sigma_{ij}=-(p+\psi) \delta_{ij}+2 \nu e_{ij}$, is the total stress tensor for an incompressible Newtonian fluid and $e_{ij}$ is the strain-rate tensor defined later. When taking the scalar product with $u_i$ and making use of the product rule, we find,
\begin{equation} \label{eq:mtm_tensor}
\frac{1}{2}\partial_t u_{i} u_i = bu_i  g_i + \partial_j(\sigma_{i j}u_i) - \sigma_{i j} \partial_j u_i. 
\end{equation}
The strain-rate tensor for an incompressible Newtonian fluid is \citep{Acheson1991}, 
\begin{equation}\label{eq:stress_tensor}
e_{i j} = \frac{1}{2} (\partial_i u_j + \partial_j u_i),
\end{equation}
therefore by noting that by definition $e_{ij}=e_{ji}$, and by extension $\sigma_{ij}=\sigma_{ji}$, using equation \ref{eq:stress_tensor} we find, 
\begin{equation}
\sigma_{i j} e_{i j} = \sigma_{ij} \partial_i u_j.
\end{equation}
Using this result in equation \ref{eq:mtm_tensor} gives,
\begin{equation}
\frac{1}{2}\partial_t u_{i} u_i = bu_i  g_i + \partial_j(\sigma_{i j}u_i) - \sigma_{i j} e_{i j}.
\end{equation}
Therefore, by expanding using the definition of $\sigma_{ij}$ in the last term,
\begin{equation}
\frac{1}{2}\partial_t u_{i} u_i = bu_i  g_i + \partial_j(\sigma_{i j}u_i) + (p+\psi)e_{i i}  -2 \nu e_{ij} e_{i j}.
\end{equation}
By noting that $e_{ii}=0$ as we are considering an incompressible system where ${\nablab \cdot \boldsymbol{u} =0}$, and by expanding all terms we find, 
\begin{equation}
\frac{1}{2}\partial_t u_{i} u_i = bu_i  g_i - \partial_j(p+\psi)\delta_{ij}u_i  -2 \nu \partial_j e_{ij}u_i-2 \nu e_{ij} e_{i j},
\end{equation}
and taking the volume integral (as in \S~\ref{sec:energy}), the following energy balance can be found, 
\begin{multline}
\dtotal{E_K}{t} = -\dtotal{E_{PE}}{t} - D_{ther} - \oint_S (p+\psi) (\delta_{ij}u_i) \hat{n}_j ~ \mathrm{dS}  \\ + 2 \nu \bigg(\oint_S e_{ij}u_i\cdot \hat{n}_j~\mathrm{dS} -\int_V ~e_{ij} e_{i j}~\mathrm{dV} \bigg).
\end{multline}
The viscous dissipation term is separated into two components; one corresponds to the dissipation within the fluid (the volume integrated component) and is as calculated in equation 31 of \cite{Ogilvie2009}. The other is the normal viscous flux through the surface (the surface integral term). 
We can now establish an additional balance in the system as it can be shown numerically that the normal viscous fluxes balance the pressure integral, and the bulk viscosity component balances the injection term (involving $\psi$) in the absence of buoyancy forces.

\section{Analytical calculation of g-modes}\label{app:g-mode}

The frequencies of the internal gravity (g-mode) resonances in a uniformly stratified medium with constant $N^2$ can be analytically calculated, given suitable approximations. To allow for straightforward analytical results we neglect viscosity and thermal diffusivity and consider solid-wall boundary conditions. Neglecting diffusion is valid if the group travel time for such (equivalent travelling) modes is significantly larger than the viscous/thermal damping timescale. 

We may reduce our system to that used in \cite{Pontin2020}, where the governing equations become
\begin{equation}
\ddtotal{\xi_r}{r} + \frac{4}{r}\dtotal{\xi_r}{r} - \bigg[ \bigg(1 - \frac{\bar{N}^2}{\omega^2}\bigg) \frac{l(l+1)}{r^2} -2 \bigg]\xi_r=0,
\end{equation}
which can be solved to give
\begin{equation}
\xi_r = A r^{\lambda_+} + B r^{\lambda_-},
\end{equation}
where  
\begin{equation}
\lambda_{\pm} = -\frac{3}{2} \pm \frac{1}{2}\sqrt{1+4\bigg(1-\frac{\bar{N}^2}{\omega^2}\bigg)l(l+1)}. 
\end{equation} 
Here we are using displacement, $\vb{\xi}=\dpartial{\vb{u}}{t}$, where $\xi_r$ denotes the radial component. 

We are considering $\bar{N}^2>0$ and therefore consider oscillatory solutions with  complex $\lambda_{\pm}$. The critical value for $\omega$  that bounds the internal gravity wave regime is found to be, 
\begin{equation}
\omega^2 < \frac{4 \bar{N}^2l (l+1)}{4 l (l+1) +1}.
\end{equation}
Defining $\lambda_i=\operatorname{Im}[\lambda]$ and $\lambda_r=\operatorname{Re}[\lambda]$, the solution for displacement $\xi_r$ can instead be written in the form, 
\begin{equation}
\xi_r = A r^{\lambda_r} e^{i \lambda_i \ln(r)} + B r^{\lambda_r} e^{-i \lambda_i \ln(r)}. 
\end{equation}
We then apply solid-wall boundary conditions at either end of the domain,
\begin{equation}
\xi_r=0 \quad r=\alpha r_0 \text{ and } r=\beta r_0.
\end{equation}
This departs from the boundary conditions used in the numerical calculations but maintains simplicity in the analytical calculations with little effect on the final result when $\omega^2 \ll \omega_d^2$. These combine to give
\begin{equation}
-e^{i \lambda_i (\ln(\beta r_0)-2\ln(\alpha r_0))} + e^{-i \lambda_i \ln(\beta r_0)}=0,
\end{equation}
from which it follows that
\begin{equation}
2 \lambda_i (\ln(\beta r_0)-\ln(\alpha r_0))= 2 \pi n,
\end{equation}
where $n$ is an integer, from $1$ to $\infty$. The final dispersion relation for the frequencies given is therefore
\begin{equation}\label{eq:g_ana}
\omega^2 = \frac{4 l (l+1) \bar{N}^2 (\ln(\beta r_0)-\ln(\alpha r_0))^2}{(2 l+1)^2 (\ln(\beta r_0)-\ln(\alpha r_0))^2+4 \pi ^2 n^2}. 
\end{equation}

\section{Analytical calculation of f-modes}\label{app:f-mode}

To quantify the variation in frequency of the surface gravity (f-mode) resonance on the parameters of the system, it is helpful to analytically compute these mode frequencies.  As in Appendix \ref{app:g-mode} we neglect the viscosity and thermal diffusivity to recover the system used in \cite{Pontin2020}.  However in this case instead we consider the equation for the pressure perturbation,
\begin{equation}
\ddtotal{p}{r} + \frac{2}{r}\dtotal{p}{r} - \bigg(1 - \frac{N^2}{\omega^2}\bigg) \frac{l(l+1)}{r^2}p=0,
\end{equation}
and include the free-surface boundary condition. This can then easily be solved to find
\begin{equation}
p = A r^{\mu_+} + B r^{\mu_-},
\end{equation}
where
\begin{equation}\label{eq:mu}
\mu_{\pm} =\lambda_{\pm} +1= -\frac{1}{2} \pm \frac{1}{2}\sqrt{1+4\bigg(1-\frac{\bar{N}^2}{\omega^2}\bigg)l(l+1)}.
\end{equation}
Using the relation, $\dtotal{p}{r}= \omega^2 \Big(1 - \frac{N^2}{\omega^2}\Big)\xi_r$, we can find the radial displacement to be
\begin{equation}
\xi_r = \frac{1}{ \omega^2 (1 - \frac{N^2}{\omega^2})} \big(\mu_+ A r^{\mu_+ -1} + \mu_- B r^{\mu_- -1} \big).
\end{equation}
We use the same boundary conditions as in the \S~\ref{sec:boundary}, no radial displacement at the core, and a free surface at the planetary radius. Therefore with $r_0=1$ these two conditions can be combined to give, 
\begin{equation}
1 - \frac{\mu_+}{\mu_-}\alpha^{\mu_+ - \mu_1} = \frac{\omega_d^2 }{\omega^2 -N^2} \Big(\mu_+ -\mu_+\alpha^{\mu_+ - \mu_-}\Big),
\end{equation}
where $\omega_d=\sqrt{\frac{GM}{r_0^3}}$. We consider this in different cases to predict the location in frequency of the f-mode peak, in each case taking the positive complex solution. 

First we consider the case with no core and no stratification, $N=0$ and $\alpha=0$, and recover the expected limit \citep{Barker2016}
\begin{equation}
\omega^2=\omega_d^2 l. 
\end{equation}
If we consider the case in which there is a stratified layer and no inner core, $N \neq 0$ and $\alpha=0$, we obtain 
\begin{equation}
\omega^2 = \frac{1}{2}\bigg(N^2 - \omega_d^2 + \sqrt{N^4 - 2N^2\omega_d^2 + (1 + 2l)^2\omega_d^4}\bigg). 
\end{equation}
Taking the limit in which there is a fully convective layer with a finite core size, $N=0$ and $\alpha \neq 0$, we find
\begin{equation}
\omega^2 = \omega_d^2 (l+1)\frac{\alpha^{-2(l+1)}-1}{\frac{l+1}{l}\alpha^{-2(l+1)}+1}. 
\end{equation}
The final case in which $N \neq 0$ and $\alpha \neq 0$ can be calculated using symbolic algebra packages (e.g. Mathematica). The frequency predictions in this case are plotted on some figures, but we do not show the expression here because it does not reduce to a simple form. 

\section{Analytical dissipation in the travelling wave regime}\label{app:trav_wave}

For a sufficiently low frequency tidal forcing $\omega^2 \ll \omega_d^2$, such as those that would be expected in some tidal applications, the forced waves will be damped before they reach the inner core and form a standing mode. In these cases therefore the energy of the waves is fully dissipated into the medium.  Note, although in this case we are considering the wave to be damped by viscosity and thermal diffusivity, this calculation is independent of the particular damping mechanism, and would apply in any circumstance in which waves are efficiently excited and then subsequently fully damped. For example, this regime would also occur if wave breaking occurs due to nonlinear effects \citep[e.g.][]{Barker2010}. We again neglect viscosity and thermal diffusivity and consider the modified pressure perturbation $W=p+\psi$ and radial displacement $\xi_r$ to obtain
\begin{eqnarray}\label{app:eqn:W}
W&=& A r^{\mu_+} + B r^{\mu_-}, \\
\label{app:eqn:xir}
\xi_r &=& \frac{1}{\rho_0 \omega^2 (1 - \frac{N^2}{\omega^2})} \big(\mu_+ A r^{\mu_+ -1} + \mu_- B r^{\mu_- -1} \big),
\end{eqnarray}
where $\mu_{\pm}$ is as defined in equation \ref{eq:mu}.

If we consider frequencies sufficiently low such that the wave is damped due to viscous forces before reflecting off the inner core and returning to the outer surface, then we can assume that at (or just inside) the outer surface only the ingoing component of the wave solution is non-zero. Hence, we assume $B=0$. We continue to use the free-surface condition used in our numerical results, 
\begin{equation}
\Delta p =0 \quad \text{at} \;\; r=r_0, 
\end{equation}
which gives us
\begin{equation}\label{app:eqn:BC2}
A= \frac{\psi r_0^{2-\mu_+} }{1  - \frac{\omega_d^2}{\rho_0  (\omega^2 - N^2)} \mu_+}.
\end{equation}

The energy flux is defined using the standard definition for a
linear wave,
\begin{equation}
F=\pi r^2 \int_0^{\pi} \operatorname{Re}[-i \omega \xi_r W^*] \sin \theta d\theta,
\end{equation}
where $\xi_r$ and $W$ have $r$ and $\theta$ dependence, which here evaluates to 
\begin{equation}
F=\frac{r^2}{2} \operatorname{Im}[\omega \xi_r W^*],
\end{equation}
where $\xi_r$ and $W$ have only $r$ dependence. Using equations \ref{app:eqn:W} and \ref{app:eqn:xir}, we calculate the flux close to the outer surface by continuing to assume $B=0$. In reality the flux is radially dependent, but we are just concerned about the flux just below the outer boundary, where $B =0$ holds if there is no reflected wave there.  Therefore the flux can be written as
\begin{equation}
F=\frac{\omega^3}{2l(l+1) \rho_0^2} \frac{|\mu_+ A|^2}{(\omega^2 - N^2)^2}  \operatorname{Im}\big[\mu_+\big]r^{0} .
\end{equation}
By substituting in $A$ and $\mu_+$ and taking the limit $\omega^2~\ll~\omega_d^2~\sim~N^2$, we obtain
\begin{equation}
F= \frac{|\psi_0|^2 r_0^{5} N \omega^2}{2 \sqrt{l(l+1)}}. 
\end{equation}
As the assumption being made is such that the wave is fully damped before reaching the inner core, we can assume that the energy dissipated is equal to the total flux of the wave, i.e. the total dissipation rate in this travelling wave calculation is given by $D_{TW}=F$. 

The frequency-averaged dissipation with a $1/\omega$ weighting, in the travelling wave dominated regime is 
\begin{equation}
\bar{D}_{TW} = \int^{\omega_{max}} \frac{D_{TW}}{\omega}~\mathrm{d} \omega = \frac{|\psi_0|^2 r_0^{5} \bar{N}}{2 \sqrt{l(l+1)}} \int^{\omega_{max}} \omega~\mathrm{d} \omega.
\end{equation} 
Therefore, it is given by 
\begin{equation}
\bar{D}_{TW}=\frac{|\psi_0|^2 r_0^{5} \bar{N} }{4 \sqrt{l(l+1)}} \Big[\omega^2 \Big]^{\omega_{max}}_{\omega_{min}}.
\end{equation}
In most instances we take $\omega_{max}=\bar{N}$, $\omega_{min}=0$, therefore
\begin{equation}
\bar{D}_{TW}=\frac{|\psi_0|^2 r_0^{5} \bar{N}^3}{4 \sqrt{l(l+1)}},
\end{equation}
but if we take $\omega_{max}=\omega_{crit}$ where $\omega_{crit}$, is defined by the limit in equation \ref{eq:omega_crit}, then
\begin{equation}
\bar{D}_{TW}=\frac{|\psi_0|^2 r_0^{5} \bar{N}^{\frac{5}{2}} (\beta-\alpha)^{\frac{1}{2}} (\nu +\kappa)^{\frac{1}{2}}  (l(l+1))^{\frac{1}{4}}}{4}.
\end{equation}

\subsection{Transition frequency} 

We are considering a wave for which its damping timescale is sufficiently short that it does not propagate back to the outer edge of the stratified layer after it has been launched. As we are considering viscous and thermal damping in our linear calculation, we can predict the frequency describing the transition point by comparing the radial group travel time for a gravity wave packet with the viscous and thermal damping timescales.

The radial group velocity is defined as
\begin{equation}
c_{g,r}=\vb{\hat{r}} \cdot \vb{c_g}= \dpartial{\omega}{k_r},
\end{equation}
where for a short wavelength plane internal gravity wave
\begin{equation}
\omega^2=\frac{k_{\perp}^2 N^2}{k_r^2+k_{\perp}^2}. 
\end{equation}
Therefore, 
\begin{equation}
c_{g,r}=\frac{-k_{\perp}k_r N}{(k_r^2+k_{\perp}^2)^{3/2}},
\end{equation}
with $k_{\perp}=\frac{\sqrt{l(l+1)}}{r_0}$. As we are considering low frequencies, we can assume that $k_{\perp} \ll k_r$, and
\begin{equation}
c_{g,r} \approx -\frac{k_{\perp}N}{k_r^2}. 
\end{equation}
Therefore, by considering that the distance required to travel is $2(\beta- \alpha)r_0$, the group travel time is
\begin{equation}
t_g=\frac{2(\beta-\alpha) r_0}{c_{g,r}}=\frac{2 (\beta-\alpha) r_0 k_r^2}{k_{\perp} N}. 
\end{equation}
The total viscous/thermal damping time is
\begin{equation}
t_{d} \approx \frac{2}{(\nu + \kappa) k_r^2},
\end{equation}
which means that taking $t_g \lesssim t_{d}$ gives
\begin{equation}
k_r^4 \lesssim\frac{N k_{\perp}}{(\beta-\alpha) r_0 (\nu+\kappa)},
\end{equation}
or equivalently in terms of our tidal forcing frequency
\begin{equation}
\omega \lesssim \big((\beta-\alpha) r_0 (\nu+\kappa) (N k_{\perp})^3 \big) ^{\frac{1}{4}}. 
\end{equation}
Therefore we expect the tidal response for frequencies smaller than this approximate value to be in the travelling wave regime in our linear calculations. In reality other effects, including wave breaking or other nonlinear effects, can cause efficient damping of propagating waves; this would perhaps alter the critical frequency but we would expect to obtain the same prediction for the dissipation rate also in this case.

\bibliographystyle{aasjournal}
\bibliography{paperII}

\end{document}